\documentclass[11pt]{article}
\usepackage[T1]{fontenc}
\usepackage[utf8]{inputenc}
\usepackage{textcomp}
\usepackage{lmodern}
\usepackage{amsmath,amssymb,bm}
\usepackage{graphicx}
\usepackage[margin=2.5cm]{geometry}
\usepackage[numbers,sort&compress]{natbib}
\usepackage[colorlinks=true,citecolor=blue,urlcolor=blue]{hyperref}
\title{Quantitative Analytical Spheroid Model for Scattering-Type Scanning Near-Field Optical Spectroscopy of Layered Samples}
\author{Kirill V. Voronin$^{1,*}$, Edoardo Vicentini$^2$, Lars Mester$^3$,\\
Rainer Hillenbrand$^{2,4,*}$, Alexey Y. Nikitin$^{1,4,*}$\\[0.7em]
\small $^1$Donostia International Physics Center (DIPC), Donostia-San Sebasti\'an 20018, Spain\\
\small $^2$CIC nanoGUNE BRTA, Tolosa Hiribidea 76, 20018 Donostia-San Sebastian, Spain\\
\small $^3$Attocube Systems GmbH, 85540 Haar, Germany\\
\small $^4$IKERBASQUE, Basque Foundation for Science, Bilbao 48013, Spain\\[0.4em]
\small $^*$Corresponding authors: kirill.voronin@dipc.org; r.hillenbrand@nanogune.eu; alexey@dipc.org}
\date{}
\begin{document}
\maketitle
\begin{abstract}
Scattering-type scanning near-field optical microscopy (s-SNOM) is a versatile tool for probing optical properties at deeply subwavelength scales. Its theoretical modeling, however, typically relies on computationally intensive numerical simulations or phenomenological analytical models, limiting quantitative spectral interpretation. In our previous work, we introduced an exact analytical solution for the quasi-electrostatic prolate spheroid model of s-SNOM for homogeneous bulk samples, combining the accuracy of numerical simulations with the efficiency of analytical approaches. Here, we generalize this framework to multilayer samples using two equivalent methods: a generalized image charge method and a Fourier integral method. The resulting exact spheroid model reproduces finite-element simulations within numerical precision while retaining the computational efficiency of an analytical model. We validate the model experimentally for poly (methyl methacrylate) layers of different thicknesses, showing that a single set of geometrical parameters simultaneously reproduces multiple demodulation harmonics and all measured thicknesses, unlike the finite dipole model. We further demonstrate its applicability to strongly resonant SiC layers on dielectric and photoexcited InAs substrates, where existing analytical approaches fail. The model provides an efficient quantitative framework for interpreting multilayer s-SNOM spectra and extracting local optical properties.
\end{abstract}

\section{Introduction}

Scattering-type scanning near-field optical microscopy (s-SNOM) is a powerful nano-optical imaging and spectroscopy technique that provides access to optical properties at spatial scales far below the diffraction limit \cite{ref1}. By employing a sharp atomic force microscope (AFM) tip to strongly concentrate the incident electromagnetic field, s-SNOM enables local near-field tip-sample interaction and nanoscale probing of a broad range of light--matter phenomena from the terahertz to the visible spectral range \cite{ref2,ref3,ref4,ref5,ref6,ref7,ref8}. In particular, s-SNOM has become an important tool for excitation and imaging of highly confined polaritons \cite{ref9,ref10}, while recent developments have extended near-field microscopy to cryogenic temperature \cite{ref11,ref12,ref13}, strong magnetic fields \cite{ref14}, and simultaneous detection of optical near fields and light-induced tunneling currents \cite{ref15,ref16,ref17}. These advances increasingly enable studies of quantum materials and quantum light at the nanoscale \cite{ref18,ref19,ref20}. The broad experimental capabilities of modern s-SNOM therefore provide access to an increasingly diverse range of nanoscale optical phenomena.

A particularly important class of systems studied by s-SNOM comprises multilayer samples. Such structures are ubiquitous in condensed-matter physics and nanophotonics \cite{ref1,ref21,ref22}, including van der Waals heterostructures \cite{ref23,ref24,ref25}, semiconductors, metallic or dielectric layers \cite{ref26,ref27}, but are also encountered in chemical \cite{ref28,ref29} and biological samples \cite{ref30,ref31}. Moreover, even structurally complex samples can often be approximated locally as laterally homogeneous layered systems within the spatial region probed by the tip \cite{ref32}. In spectroscopic measurements, the near-field response of such structures contains information about the local dielectric properties, layer thicknesses, and electromagnetic coupling between different constituents of the sample. A quantitative model connecting these properties to the measured near-field spectra is therefore essential for extracting material parameters at deeply subwavelength scales and, ultimately, for developing s-SNOM toward quantitative nanoscale optical tomography \cite{ref33,ref34,ref35}.

Despite the remarkable progress of experimental near-field microscopy, quantitative interpretation of s-SNOM spectra remains considerably less developed than the experimental capabilities of the technique. Rigorous computational approaches based on finite-element (FEM), finite-difference time-domain (FDTD), or boundary-element (BEM) methods \cite{ref26,ref33,ref36,ref37,ref38} can accurately describe the tip--sample interaction, but are very computationally expensive \cite{ref39,ref40,ref41,ref42,ref43}. Since the exact geometry of an experimental s-SNOM tip is generally not known, numerical simulations usually rely on idealized tip geometries. A widely used approximation is an elongated prolate spheroid, which captures the finite extent of the tip and the realistic tip apex shape \cite{ref26,ref37,ref38}. Numerical spheroidal-tip models have been successfully applied to both layered samples and strongly resonant tip--sample interactions \cite{ref26,ref37}. Analytical models, such as the point-dipole model (PDM) \cite{ref44}, finite-dipole model (FDM) \cite{ref45}, or more advanced analogs \cite{ref46}, provide a much faster alternative, but rely on simplified descriptions of the tip, usually based on the spheroid model, but using a set of discrete effective charges instead of a continuous charge distribution at the tip surface. For homogeneous bulk samples, an appropriately fitted FDM can reproduce experimental spectra quantitatively \cite{ref45}. The original FDM was formulated for semi-infinite samples, for which the sample response can be described by a momentum-independent electrostatic reflection coefficient. It was subsequently generalized to layered samples using additional approximations to incorporate their momentum-dependent response. Namely, Hauer et al. \cite{ref47} represented the response of the multilayer to each effective charge representing the tip in the FDM by a single effective image charge, whose strength and position are chosen to reproduce the electrostatic potential and normal electric field at the sample surface. An alternative approach introduced by Mester et al. \cite{ref48} replaces the momentum-dependent reflection coefficient of the multilayer with an effective coefficient obtained by averaging over in-plane momenta using a weight function determined by the position of a representative charge near the tip. Both approaches were developed and validated for weakly resonant layers, in particular molecular vibrational resonances, and are therefore intended primarily for weak oscillators rather than strongly resonant polaritonic systems \cite{ref47,ref48}. Both multilayer extensions are implemented in the open-source code \cite{ref49}.

Although these multilayer extensions provide computationally efficient descriptions of weakly resonant films, their approximations have known limitations outside this regime. Hauer et al. \cite{ref47} found the largest deviations of their effective-image-charge approximation when the tip couples resonantly to a thin film on a dielectric substrate and recommended explicitly verifying the approximation in such cases. More generally, comparisons of the PDM and the FDM with rigorous numerical calculations have revealed substantial discrepancies for strongly resonant materials such as SiO\textsubscript{2}, Al\textsubscript{2}O\textsubscript{3}, and SiC \cite{ref37}, where several probe--sample modes can contribute to the near-field response. These observations motivate a more systematic assessment of analytical models for layered samples. In particular, for quantitative spectroscopy it is desirable that model parameters associated with the tip--sample geometry remain transferable between measurements in which only the sample thickness or demodulation harmonic number is changed. As we demonstrate below, the commonly used multilayer FDM of Hauer et al. does not provide such a simultaneous description for the experimental datasets considered here: for weakly resonant poly (methyl methacrylate) PMMA layers, a single parameter set does not quantitatively reproduce different thicknesses and demodulation harmonics, while for strongly resonant SiC layers additional limitations in spectral shape and convergence become apparent.

As a step toward a rigorous and computationally efficient description of s-SNOM, in our previous work we derived an exact analytical solution of the quasi-electrostatic prolate spheroid model for homogeneous bulk samples \cite{ref50}. The model determines the polarization-charge distribution on the spheroidal tip and reproduces finite-element calculations within numerical precision while retaining the computational efficiency characteristic of analytical models. Importantly, the solution was formulated such that the interaction of the spheroid with the semi-infinite sample is described through electrostatic image charges which allows the extension to more complex nanostructures. Hereinafter, for brevity, we call this approach the exact spheroid model (ESM).

Here, we extend the ESM to planar multilayer structures. The key step is incorporating the electrostatic response of the multilayer sample to the ESM. To do this, we develop two alternative methods: a generalized image-charge method (GICM), which represents the wavevector-dependent electrostatic response of the multilayer by a finite series of effective image charges, and the Fourier integral method (FIM), which represents the sample response as an integral over in-plane Fourier harmonics. Both solutions modify only the description of the sample response within the ESM; the spheroidal-tip model and the overall solution procedure remain unchanged. Both methods give the same result and have the same order of computational complexity, but their relative efficiency depends on the conditions of the problem. The resulting multilayer ESM (MESM) can be systematically converged to arbitrary accuracy while remaining only moderately more computationally demanding than the bulk-sample formulation. The approach is not limited to a single layer on a substrate; it applies more generally to laterally homogeneous multilayer structures. We benchmark the resulting model experimentally using PMMA layers of different thicknesses on Si and find quantitative agreement with the measured spectra for all layer thicknesses and demodulation harmonics using the same geometrical parameters. We further apply the model to a strongly resonant SiC layer on photoexcitable InAs and reproduce the near-field spectra of both the non-excited and photoexcited structures with the same parameters, including the pronounced spectral changes induced by photoexcitation.

\section{Results and discussions}

\subsection{Theoretical basis}

Scattering-type scanning near-field optical microscopy is based on AFM, in which a sharp vertically oscillating tip is illuminated by focused electromagnetic radiation, as schematically illustrated in Figure 1a. The incident field induces polarization charges in the tip, while the near-field interaction with the sample modifies their distribution. Consequently, the amplitude and phase of the light scattered by the tip contain information about the local optical response of the sample. To suppress the far-field background, the interferometrically detected signal is demodulated at higher harmonics of the tip oscillation frequency, typically \(n = 2,\ 3,\ 4\). Because these higher harmonics originate predominantly from the strongly nonlinear distance dependence of the near-field interaction close to the tip apex, the apex geometry plays the dominant role in determining the measured near-field response.

To model the tip-sample interaction, we approximate the tip by a prolate spheroid of length \(L = 2a\) with major semi-axis \(a\) and minor semi-axis \(b\), which are related to the tip apex curvature radius as \(R = \frac{b^{2}}{a}\), placed at the distance \(H\) above the sample surface (Figure 1b), which oscillates as \(H(t) = H_{0} + A(1 - \cos{\Omega t})\), where \(A\) and \(\Omega\) are the tapping amplitude and frequency, respectively. Since the tip curvature radius and tapping amplitude are much smaller than the illumination wavelength in the spectral range considered here, we employ the quasi-electrostatic approximation and neglect retardation and magnetic-field effects. Furthermore, because of the elongated tip geometry, we consider only the polarization along the spheroid major axis and the vertical component of the incident field. Within these approximations, the scattered field can be expressed as \cite{ref32,ref45}

\begin{equation}
E_{\mathrm{sca}}(H) \propto \left( 1 + \varsigma r_{\mathrm{p}} \right)^{2}\alpha_{z}(H)E_{\mathrm{inc},z},
\label{eq:1}
\end{equation}

\noindent where \(\alpha_{z}(H)\) is the vertical polarizability of the spheroid, \(r_{\mathrm{p}}\) is the Fresnel reflection coefficient for p-polarized light, and \(\varsigma\) accounts for far-field reflection effects \cite{ref32}; throughout this work we use \(\varsigma = 1\). The experimentally detected n\textsuperscript{th}-harmonic signal is extracted via Fourier decomposition \cite{ref50}:

\begin{equation}
S_{n} \propto \Omega\int\limits_{0}^{\frac{2\pi}{\Omega}}{E_{\mathrm{sca}}(H)e^{- in\Omega t}dt} \propto \left( 1 + \varsigma r_{\mathrm{p}} \right)^{2}\alpha_{nz}(H)E_{\mathrm{inc},z}.
\label{eq:2}
\end{equation}

In experiment, the near-field signal is typically normalized to that obtained from a reference material, yielding \cite{ref1}

\begin{equation}
\sigma_{n} = \frac{S_{\mathrm{s},n}}{S_{\mathrm{r},n}} = \frac{\left( 1 + \varsigma r_{\mathrm{p},\mathrm{s}} \right)^{2}\alpha_{\mathrm{s},n,z}}{\left( 1 + \varsigma r_{\mathrm{p},\mathrm{r}} \right)^{2}\alpha_{\mathrm{r},n,z}},
\label{eq:3}
\end{equation}

\noindent where \(S_{\mathrm{s},n}\), \(r_{\mathrm{p},\mathrm{s}}\), \(\alpha_{\mathrm{s},n,z}\) are the near-field signal, Fresnel reflection coefficient, and n\textsuperscript{th} harmonic of the spheroid polarizability of the sample, respectively, and \(S_{\mathrm{r},n}\), \(r_{\mathrm{p},\mathrm{r}}\), \(\alpha_{\mathrm{r},n,z}\) are those of the reference material. The polarizability is determined by the polarization-charge distribution induced at the spheroid surface by the external field and by its electrostatic interaction with the sample (Figure 1b).

In our previous work, we obtained an exact solution of this problem for a homogeneous bulk sample by solving Laplace\textquotesingle s equation in spheroidal coordinates with continuity of the electrostatic potential and of the normal component of the electric displacement at the tip and sample interfaces. To address the boundary condition problem at the sample surface we applied the electrostatic image method \cite{ref51}, which significantly simplify the calculations. The resulting polarizability can be written as \cite{ref50}

\begin{equation}
\alpha_{z}(H) = 2ac\vartheta_{1}(H) + \alpha_{z0},
\label{eq:4}
\end{equation}

\noindent where \(c = \sqrt{a^{2} - b^{2}}\), \(\alpha_{z0}\) is the free-space spheroid polarizability, which does not contribute to the demodulated signal, and \(\vartheta_{1}(H)\) is obtained from the infinite linear system

\begin{equation}
\sum_{n = 1}^{\infty}\mathcal{M}_{mn}\vartheta_{n} = \mathcal{C}_{m},
\label{eq:5}
\end{equation}

which is in practice truncated at a sufficiently large order \(N\), chosen from the convergence of the calculated polarizability. This system is derived by applying boundary conditions on the surface of a spheroid, taking into account the multipole expansion of the field and charge on its surface, as shown in Section S1 of the Supporting Information. The coefficients \(\mathcal{M}_{mn}\) and \(\mathcal{C}_{m}\) read

\begin{equation}
\mathcal{M}_{mn} = \delta_{mn} - \frac{\beta(\varepsilon - 1)(2n + 1)J_{nm}}{2P_{m}(\xi_{0})\left( \varepsilon\frac{Q_{m}(\xi_{0})}{P_{m}(\xi_{0})} - \frac{\xi_{0}Q_{m}\left( \xi_{0} \right) - Q_{m - 1}(\xi_{0})}{\xi_{0}P_{m}\left( \xi_{0} \right) - P_{m - 1}(\xi_{0})} \right)},
\label{eq:6}
\end{equation}

\begin{equation}
\mathcal{C}_{m} = - \frac{\beta c(\varepsilon - 1)^{2}J_{1m}}{4P_{m}\left( \xi_{0} \right)\left( \varepsilon Q_{1}\left( \xi_{0} \right) - \xi_{0}Q_{0}\left( \xi_{0} \right) + \frac{\xi_{0}^{2}}{\xi_{0}^{2} - 1} \right)\left( \varepsilon\frac{Q_{m}\left( \xi_{0} \right)}{P_{m}\left( \xi_{0} \right)} - \frac{\xi_{0}Q_{m}\left( \xi_{0} \right) - Q_{m - 1}\left( \xi_{0} \right)}{\xi_{0}P_{m}\left( \xi_{0} \right) - P_{m - 1}\left( \xi_{0} \right)} \right)},
\label{eq:7}
\end{equation}

\noindent where \(\varepsilon\) is the dielectric permittivity of the tip, \(P_{n}\) and \(Q_{n}\) are Legendre functions of the first and second kind, respectively, \(\xi_{0} = a/c\), \(\beta = \frac{\varepsilon_{\mathrm{s}} - 1}{\varepsilon_{\mathrm{s}} + 1}\), and \(\varepsilon_{\mathrm{s}}\) is the dielectric permittivity of the sample.

These coefficients depend on the following integrals, which are determined by the geometry of the system and are independent of its optical properties, and therefore need to be calculated only once:

\begin{equation}
J_{nm} = \int\limits_{- 1}^{1}{d\eta P_{n}(\eta)P_{m}(\eta_{q})Q_{m}(\xi_{q})},
\label{eq:8}
\end{equation}

\noindent where \(\left( \xi_{q},\ \eta_{q} \right)\) are the spheroidal coordinates of the image charges induced in the sample in a coordinate system associated with the original spheroid. In the following we will present an analytical expression for these integrals. Thus, \(J_{nm}\) describes the electrostatic coupling between the real polarization charge at the spheroid surface and its image in the sample provided by the electrostatic image-charge method (see Section S1 of the Supporting Information for details).

Thus, the image-charge method is a key to the analytical solution of the spheroid problem because it allows the electrostatic response of the sample to be represented by discrete image charges, avoiding explicit integration over the continuous polarization-charge distribution induced at the sample surface (see Section S2, Supporting Information and \cite{ref50} for more details). To derive the image-charge method, we consider a point charge \(q\) located at \(z_{0}\) above a planar, laterally homogeneous and in-plane isotropic sample, as illustrated in Figure 1c. The electrostatic potential in the upper half-space can be written as

\begin{equation}
\varphi\left( \mathbf{r},z \right) = q\left( \frac{1}{\sqrt{\left| \mathbf{r -}\mathbf{r}_{0} \right|^{2} + \left( z - z_{0} \right)^{2}}} + \iint\limits_{\mathbb{R}^{2}}^{}{\frac{d^{2}\mathbf{k}_{\parallel}}{2\pi k_{\parallel}}r_{\mathrm{e}}\left( \mathbf{k}_{\parallel} \right)e^{i\mathbf{k}_{\parallel}\left( \mathbf{r -}\mathbf{r}_{0} \right) - k_{\parallel}\left( z + z_{0} \right)}} \right),
\label{eq:9}
\end{equation}

\noindent where \(\mathbf{r}\) and \(z\) are the in-plane and out-of-plane coordinates, respectively, (\(\mathbf{r}_{0}\), \(z_{0}\)) are the charge coordinates, \(\mathbf{k}_{\parallel}\) is the in-plane wavevector, defining the spatial Fourier harmonics of the potential, \(r_{\mathrm{e}}\left( \mathbf{k}_{\parallel} \right)\) is the electrostatic reflection coefficient of the sample, and \(\mathbb{R}^{2}\) is the two-dimensional real plane. The first term in Equation (9) represents the potential of the source charge, whereas the second term describes the electrostatic response of the sample.

For a homogeneous bulk material with dielectric permittivity \(\varepsilon\), the electrostatic reflection coefficient is independent of the in-plane wavevector,

\begin{equation}
r_{\mathrm{e}}\left( \mathbf{k}_{\parallel} \right) \equiv - \beta_{0} = \frac{1 - \varepsilon}{1 + \varepsilon},
\label{eq:10}
\end{equation}

and the integral in Equation (7) can therefore be evaluated analytically:

\begin{equation}
\varphi\left( \mathbf{r},z \right) = q\left( \frac{1}{\sqrt{\left| \mathbf{r -}\mathbf{r}_{0} \right|^{2} + \left( z - z_{0} \right)^{2}}} - \frac{\beta_{0}}{\sqrt{\left| \mathbf{r -}\mathbf{r}_{0} \right|^{2} + \left( z + z_{0} \right)^{2}}} \right),
\label{eq:11}
\end{equation}

The second term in Equation (11) is identical to the potential produced by an image charge \(q_{\mathrm{i}} = - \beta q\) located at \((\mathbf{r}_{0},\  - z_{0})\). Thus, the electrostatic response of a homogeneous bulk sample to an arbitrary source charge can be represented by a single image charge. This property underlies the analytical solution of the spheroid model discussed above.

The situation becomes more complex for a layered sample. Consider first a single isotropic layer of thickness \(d\) and dielectric permittivity \(\varepsilon_{\mathrm{l}}\), placed on a semi-infinite substrate with dielectric permittivity \(\varepsilon_{\mathrm{s}}\). Its electrostatic reflection coefficient is wavevector dependent and reads

\begin{equation}
r_{\mathrm{e}}\left( k_{\parallel} \right) = - \frac{\beta_{1} + \beta_{2}e^{- 2k_{\parallel}d}}{1 + \beta_{1}\beta_{2}e^{- 2k_{\parallel}d}},
\label{eq:12}
\end{equation}

\noindent where the coefficients \(\beta_{1}\) and \(\beta_{2}\) are given by the following

\begin{equation}
\beta_{1} = \frac{\varepsilon_{\mathrm{l}} - 1}{\varepsilon_{\mathrm{l}} + 1},\ \ \beta_{2} = \frac{\varepsilon_{\mathrm{s}} - \varepsilon_{\mathrm{l}}}{\varepsilon_{\mathrm{s}} + \varepsilon_{\mathrm{l}}}.
\label{eq:13}
\end{equation}

If \(\left| \beta_{1}\beta_{2} \right| < 1\), this wavevector-dependent reflection coefficient can be expanded into Taylor series over \(\beta_{1}\beta_{2}e^{- 2k_{\parallel}d}\), corresponding to a conventional image-charge series \cite{ref52}, so that the electrostatic response of the layer is represented by an infinite sequence of image charges beneath the surface (see Section S2 of the Supporting Information for details). This expansion, however, converges only when \(\left| \beta_{1}\beta_{2} \right| < 1\), a condition that can readily be violated for strongly resonant layers, particularly in spectral regions where the real part of the layer permittivity becomes negative.

To overcome this restriction, we introduce the GICM. We note that for a material with finite optical losses, the pole associated with the denominator of Equation (12) is shifted away from the real axis, and the reflection coefficient is a finite and smooth function of the variable \(x = e^{- 2k_{\parallel}d}\), defined on the interval \(x \in \lbrack 0,\ \ 1\rbrack\). According to the Weierstrass approximation theorem, such a function can be approximated to arbitrary accuracy by a polynomial in \(x\) (see Section S2 of the Supporting Information for details). We therefore represent the reflection coefficient as

\begin{equation}
r_{\mathrm{e}}\left( k_{\parallel} \right) \approx - \beta_{1} - \sum_{n = 1}^{K}{C_{n}e^{- 2nk_{\parallel}d}}.
\label{eq:14}
\end{equation}

The approximation error can be made arbitrarily small by increasing the polynomial order \(K\). The coefficients \(C_{n}\) can be obtained, for example, by least-squares fitting of the function \(r_{\mathrm{e}}\left( k_{\parallel} \right)\). More generally, \(r_{\mathrm{e}}\left( k_{\parallel} \right)\) can be expanded in powers of \(e^{- 2k_{\parallel}d_{\mathrm{eff}}}\) for an arbitrary \(d_{\mathrm{eff}}\). The choice of \(d_{\mathrm{eff}}\) affects the convergence rate, i.e., the number of terms \(K\) required to reach a specified accuracy. We analyze this dependence in Section S2 of the Supporting Information and find that, for the single-layer problem considered here, \(d_{\mathrm{eff}} \leq d\) provides the most efficient representation.

The representation in Equation (14) again allows the integral in Equation (9) to be evaluated analytically, giving

\begin{equation}
\varphi\left( \mathbf{r},z \right) = q\left( \frac{1}{\sqrt{\left| \mathbf{r -}\mathbf{r}_{0} \right|^{2} + \left( z - z_{0} \right)^{2}}} - \frac{\beta_{1}}{\sqrt{\left| \mathbf{r -}\mathbf{r}_{0} \right|^{2} + \left( z + z_{0} \right)^{2}}} - \sum_{n = 1}^{K}\frac{C_{n}}{\sqrt{\left| \mathbf{r -}\mathbf{r}_{0} \right|^{2} + \left( z + z_{0} + 2nd \right)^{2}}} \right).
\label{eq:15}
\end{equation}

Thus, within the GICM, the electrostatic response of a single layer on a semi-infinite substrate (Figure 1c) is represented by the image charge \(q_{0} = - \beta_{1}q\) at a depth \(z_{0}\), together with \(K\) additional image charges \(q_{n} = - C_{n}q\) located at depths \(z_{n} = z_{0} + 2nd\), where \(n\  = 1,\ 2,\ldots,K\). Unlike the conventional series, this representation is not restricted by the condition \(\left| \beta_{1}\beta_{2} \right| < 1\), and its accuracy can be systematically increased by increasing \(K\).

We now return to the spheroid solution introduced in Equations (4)--(8) to show explicitly how the multilayer response enters the ESM. For a homogeneous sample, the spheroid polarizability \(\alpha_{z}(H)\) follows directly from \(\vartheta_{1}(H)\) via Equation (4), while the coefficients \(\vartheta_{n}(H)\) are determined by the linear system in Equation (5). The sample response enters the coefficients \(\mathcal{M}_{mn}\) and \(C_{m}\) of this system through the terms \(\beta J_{nm}\), where the integrals \(J_{nm}\), defined in Equation (8), depend only on the tip geometry and the distance between the tip and its image (see Section S1 of the Supporting Information).

For a layered sample, the GICM derived above adds to the single image charge the set of \(K\) additional image charges. Accordingly, as shown in Section S3 of the Supporting Information, in Equations (5) and (6), every occurrence of \(\beta J_{nm}\) should be replaced as follows

\begin{equation}
\beta J_{nm} \rightarrow \beta_{1}J_{nm} + \sum_{k = 1}^{K}{C_{k}J_{nm}^{k}}.
\label{eq:16}
\end{equation}

Here, \(J_{nm}^{k}\) has exactly the same form as the integral in Equation (8), but is evaluated using the spheroidal coordinates \(\left( \xi_{q}^{k},\eta_{q}^{k} \right)\) of the k\textsuperscript{th} effective image charge. The complete expressions for the modified \(\mathcal{M}_{mn}\) and \(C_{m}\) are given in Section S3 of the Supporting Information. Thus, the multilayer extension changes only the representation of the sample response, while the structure and solution procedure of the ESM remain unchanged.

An alternative way to implement the MESM is to keep integral in Equation (9) and directly substitute this expression for the sample potential to the boundary condition at the spheroid surface. This approach called the FIM is derived in Section S4 of the Supporting Information \cite{ref53,ref54}. The multilayer response calculated using the FIM enters the ESM in a similar way to the GICM, namely, in Equations (5) and (6), every occurrence of \(\beta J_{nm}\) should be replaced as follows

\begin{equation}
\beta J_{nm} \rightarrow \mathcal{J}_{nm} = - \pi P_{n}\left( \xi_{0} \right)\int\limits_{0}^{\infty}\frac{r_{\mathrm{e}}\left( k_{\parallel} \right)}{k_{\parallel}}I_{n + \frac{1}{2}}\left( k_{\parallel}c \right)I_{m + \frac{1}{2}}\left( k_{\parallel}c \right)e^{- 2k_{\parallel}(a + H)}dk_{\parallel}.
\label{eq:17}
\end{equation}

\noindent where \(I_{\nu}\) is the modified Bessel function of the first kind.

Thus, the difference between the outcomes of the GICM and the FIM, given by Equations (16) and (17), is that the first method sums over the image charges. In contrast, the second one requires integration over the in-plane momentum, which can be evaluated numerically as a finite sum using a quadrature method. According to our convergence tests, the required number of image charges (5-20) is typically smaller than the required number of points over momenta needed for the numerical integration (20-40). Nevertheless, both methods, when implemented with numerical evaluation of the finite series, have the same order of computational efficiency, and the relative efficiency depends on each method\textquotesingle s implementation and optimization quality. In this work, we use the GICM unless otherwise stated because we find it more physical, intuitive, and much easier to implement in code than the FIM. However, as we show below, for some specific cases, such as hyperbolic uniaxial samples (e.g., hBN), the FIM is more applicable because the GICM experiences convergence issues.

Finally, we emphasize that the FIM and the GICM are not different models, but two equivalent approaches to solving the boundary-condition problem on the surface of a sample, with parallels that can be drawn at any stage of the solution. In particular, substituting Equation (14), which captures the essence of the GICM, into Equation (17) yields an equation equivalent to Equation (16), providing an alternative expression for the integrals \(J_{nm}\) and \(J_{nm}^{k}\). Importantly, this alternative integral representation allows us to find an analytical expression for these integrals in terms of the hypergeometric function, eliminating numerical integration from our solution for the bulk ESM \cite{ref50} and the GICM-based implementation of the MESM. We present these calculations and expressions in Section S5 of the Supporting Information.

\subsection{Calculation results}

To validate the accuracy of the MESM, we compare its predictions with FEM simulations of the same tip--sample geometry. Following the approach used for the bulk-sample model \cite{ref50}, we consider a hypothetical material whose dielectric function combines representative optical responses encountered in s-SNOM experiments. Specifically, the permittivity contains a Drude contribution (\(\omega_{\mathrm{p}} = 500\,\mathrm{cm}^{-1}\), \(\gamma_{\mathrm{p}} = 150\,\mathrm{cm}^{-1}\), and \(\varepsilon_{\infty} = 5\)), a strong Lorentz resonance (\(A_{\mathrm{S}} = 1200\,\mathrm{cm}^{-1}\), \(\omega_{\mathrm{S}} = 1000\,\mathrm{cm}^{-1}\), and \(\gamma_{\mathrm{S}} = 60\,\mathrm{cm}^{-1}\)), and a weak Lorentz resonance (\(A_{\mathrm{w}} = 500\,\mathrm{cm}^{-1}\), \(\omega_{\mathrm{w}} = 1800\,\mathrm{cm}^{-1}\), and \(\gamma_{\mathrm{w}} = 30\,\mathrm{cm}^{-1}\)), as shown in Figure 1d. These contributions represent, respectively, free-carrier responses characteristic of metals and doped semiconductors, strong resonances such as optical phonons or excitons, and weak molecular vibrational resonances. The hypothetical material forms a 20-nm-thick layer on a Si substrate with \(\varepsilon_{\mathrm{Si}} = 12\). The spheroidal tip is described by a curvature radius \(R = 30\,\mathrm{nm}\), total length \(L = 600\,\mathrm{nm}\), tapping amplitude \(A = 50\,\mathrm{nm}\), and minimum tip--sample distance \(H_{0} = 3\,\mathrm{nm}\).

Figure 1e shows the amplitude and phase of the second, third, and fourth harmonics of the demodulated spheroid polarizability above the layered sample, normalized to those calculated above bulk Si. The solid curves represent the MESM calculations, whereas the dashed curves correspond to the FEM simulations described in Section S6 of the Supporting Information. The overall spectral response resembles that obtained for a bulk material with the same dielectric function in our previous work \cite{ref50}, but the finite layer thickness introduces additional spectral features. In particular, the near-field spectrum peaks corresponding to both the Drude and the strong Lorentz resonances split into two distinct peaks. This splitting originates from hybridization of the surface modes supported at the upper and lower interfaces of the thin layer, producing two coupled thin-film polariton branches. For a symmetric free-standing polar film, these branches correspond to the even and odd modes with respect to the out-of-plane electric field \cite{ref55}. The effect is most pronounced for the Drude and strong Lorentz resonances, for which the surface response is sufficiently strong, whereas the weak Lorentz resonance remains much closer to the bulk-like spectral shape. Despite this substantially richer multilayer response, the analytical and FEM calculations exhibit excellent agreement over the entire spectral range and for all three demodulation harmonics. The agreement within numerical precision confirms that representing the wavevector-dependent response of the layered sample by the series of image charges does not introduce a measurable loss of accuracy and that the resulting model reproduces the direct numerical solution for the corresponding geometry. Good convergence in this case is achieved with only \(K = 5\) charge-images.

\begin{figure}[!t]
\centering
\includegraphics[width=\linewidth,height=0.46\textheight,keepaspectratio]{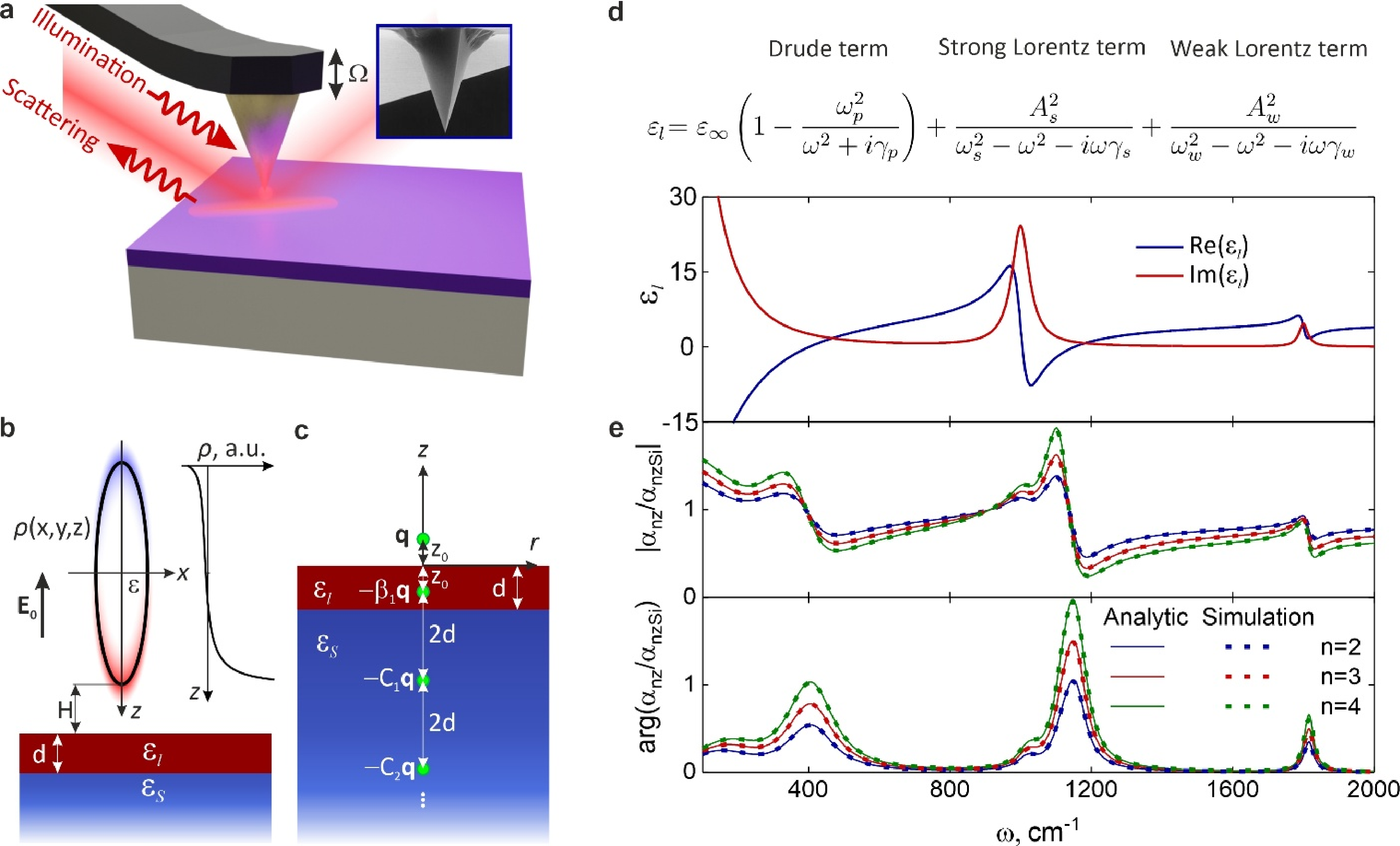}
\caption{a) Schematic of an s-SNOM cantilever above a sample consisting of a thin layer on a semi-infinite substrate. Inset: SEM image of a typical s-SNOM tip provided by NanoWorld. b) Prolate spheroid placed above the layered sample in a uniform vertically polarized electric field. The color scale represents the induced surface charge density on the spheroid, with red and blue corresponding to positive and negative charge, respectively. Inset: surface charge density as a function of the \(z\)-coordinate along the spheroid. c) Schematic illustration of the GICM, in which the electrostatic response of the layered sample to a source charge \(q\) is represented by a series of image charges below the sample surface. d) Dielectric permittivity of the hypothetical material layer, comprising a Drude contribution, a strong Lorentz resonance, and a weak Lorentz resonance. e) Amplitude and phase of the second, third, and fourth harmonics of the demodulated spheroid polarizability above a 20-nm-thick model layer on a Si substrate, normalized to the corresponding response above bulk Si. Solid curves show the analytical solution provided by the MESM, and dashed curves show FEM simulations. The geometrical parameters are \(R = 30\,\mathrm{nm}\), \(L = 600\,\mathrm{nm}\), \(A = 50\,\mathrm{nm}\), \(H_{0} = 3\,\mathrm{nm}\).}
\label{fig:1}
\end{figure}

To better understand the applicability and convergence of the MESM based on the GICM, we perform additional calculations for layers with different dielectric responses and thicknesses (Section S7, Supporting Information). These tests establish when a given number of image charges is sufficient and when a larger expansion order is required. Two cases require particular attention. First, for layers supporting highly confined, low-loss modes, the electrostatic reflection coefficient develops a sharp feature at an in-plane wavevector close to the polaritonic wavevector. Although this response can still be represented by the polynomial expansion introduced in Equation (14), a higher polynomial order is generally required, corresponding to a larger number of image charges needed to accurately reproduce the field induced in the layer. Second, thinner layers also require a larger number of image charges for convergence. This behavior is physically intuitive: as the layer thickness decreases, successive image charges are positioned more closely beneath the sample surface, and their interaction with the source charge therefore decays more slowly with increasing image-charge order. Overall, for most practically relevant cases, convergence is achieved at \(K = 10\) charges. In particularly demanding cases, a larger \(K\) may be required; for example, to calculate the spectra of thin SiC layers near its photon resonance, \(K = 20\) is required.

Importantly, the GICM is not restricted to a single layer on a semi-infinite substrate. The method can be applied to any laterally homogeneous, isotropic multilayer structure whose electrostatic response is described by a wavevector-dependent reflection coefficient. In this case, the same approximation introduced in Equation (15) remains applicable. For efficient convergence, we recommend choosing the spacing between successive image charges equal to the thickness of the thinnest layer in the stack. To demonstrate the generality of the method, we additionally compare the analytical results with FEM simulations for a SiO\textsubscript{2}/SiC/Si multilayer; the corresponding results are presented in Section S8 of the Supporting Information. These tests show that the GICM enables the ESM to treat planar multilayer samples efficiently.

Lastly, we test the MESM on a multilayer sample with uniaxial hyperbolic layers. Specifically, we analyzed a structure consisting of two 10-nm-thick hBN layers separated by a 20-nm-thick Si layer on a bulk Si substrate. Because the hBN layer supports a high, theoretically infinite, number of polaritonic modes \cite{ref56,ref57}, the GICM converges slowly and requires many more charges to describe the rapidly oscillating reflection coefficient. As a result, the GICM-based MESM become unstable and do not converge well. Consequently, to describe this type of structure, we implemented the FIM using Equation (17), which allows us to exactly reproduce the results of numerical simulations as shown in Section S9 of the Supporting Information. In summary, both the GICM and the FIM apply to arbitrary multilayer structures, but for uniaxial layers with an axis orthogonal to the sample surface, the FIM is significantly more efficient.

\subsection{Experimental validation}

To experimentally validate the ESM for layered samples, we use it to model the experimentally measured near-field spectra of PMMA layers with different thicknesses on a Si substrate, reported by Mester et al. \cite{ref48}. Figure 2a schematically illustrates the geometry used for the MESM calculations, while Figure 2b shows the dielectric permittivity of PMMA, taken from \cite{ref48}. Figures 2c and 2d compare the measured and calculated amplitude and phase of the normalized third- and fourth-harmonic near-field signals, \(s_{n} = |\sigma_{n}|\) and \(\varphi_{n} = arg(\sigma_{n})\), \(n = 3,4\), respectively, for PMMA thicknesses of 16, 40, and 100 nm. The spectra are normalized to the near-field response of bulk Si. In the calculations, we used the nominal tip radius of curvature, \(R = 30\,\mathrm{nm}\), and the tapping amplitude of \(A = 40\,\mathrm{nm}\). The spheroid length \(L\) is the only model parameter that cannot be directly associated with the tip geometry. As is known, near-field spectra of weakly resonant dielectric samples are not very sensitive to this parameter, and values around several hundred nanometers agree well with the experiments \cite{ref50}. We found that spheroid lengths from 300 nm to 500 nm work well for PMMA layers; consequently, we choose \(L = 400\,\mathrm{nm}\). The minimum tip--sample distance, \(H_{0}\), however, cannot generally be determined with nanometer precision in a conventional s-SNOM experiment because the effective separation is influenced by tip--sample forces, surface roughness, contamination layers, and other poorly controlled factors. We therefore treat the minimum distances above PMMA and the Si reference as fitting parameters, obtaining \(H_{0\mathrm{PMMA}} = 3\,\mathrm{nm}\) and \(H_{0\mathrm{Si}} = 4.2\,\mathrm{nm}\), both within the physically reasonable of 1--5 nm \cite{ref50}.

Most importantly, the same set of geometrical parameters is used for all three PMMA thicknesses and both demodulation harmonics. The MESM reproduces the evolution of both the amplitude, \(s_{n}\), and phase, \(\varphi_{n}\), spectra with layer thickness, including the changes in the spectral line shape and signal magnitude, without readjusting the model parameters for each spectrum (Figures 2c,d). This simultaneous agreement is considerably more restrictive than fitting an individual near-field spectrum: changing the layer thickness modifies the momentum-dependent response of the layered sample, while changing the demodulation order probes a different nonlinear component of the tip--sample interaction. Thus, reproducing all these observables with one parameter, which is not possible with the multilayer FDM, as shown in Figures 2e,f, provides a stringent experimental test of the multilayer model.

\begin{figure}[!t]
\centering
\includegraphics[width=\linewidth,height=0.49\textheight,keepaspectratio]{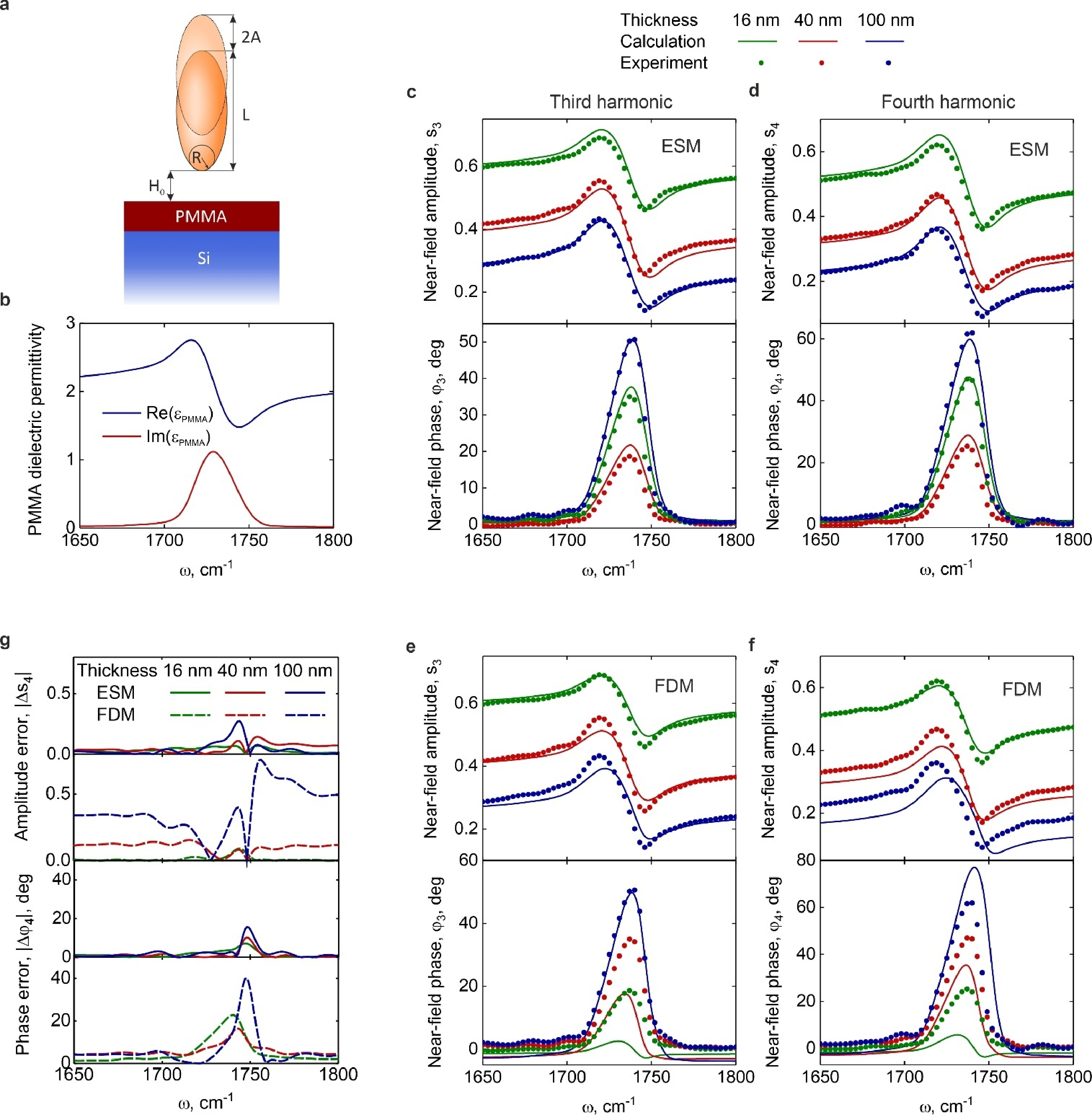}
\caption{a) Schematic of the oscillating prolate spheroid above a PMMA layer on a Si substrate. The geometrical parameters of the model are \(R = 30\,\mathrm{nm}\), \(A = 40\,\mathrm{nm}\), \(L = 400\,\mathrm{nm}\); the minimum tip--sample distance above PMMA is \(H_{0\mathrm{PMMA}} = 3\,\mathrm{nm}\). The near-field signal is normalized to that of bulk Si, for which the minimum tip--sample distance is \(H_{0\mathrm{Si}} = 4.2\,\mathrm{nm}\) and the dielectric permittivity is \(\varepsilon_{\mathrm{Si}} = 11.7\). b) Dielectric permittivity of PMMA as a function of incident-light frequency, taken from \cite{ref48}. c,d) Amplitude and phase of the normalized third- (c) and fourth-harmonic (d) near-field spectra of PMMA layers with thicknesses of 16, 40, and 100 nm. Solid curves show spectra calculated using the MESM, and symbols represent the experimental data. e,f) Corresponding third- (e) and fourth-harmonic (f) spectra calculated using the FDM with the same geometrical parameters used for MESM and \(g = 0.76e^{0.07i}\). g) The modulus of the difference between the measured amplitude (top) and phase (bottom) of the near-field signal fourth harmonic and those calculated using the MESM (solid curves) and the FDM (dashed curves) for the three PMMA thicknesses.}
\label{fig:2}
\end{figure}

To demonstrate this advantage, we compare the MESM with the FDM extension for multilayer samples developed by Hauer et al. \cite{ref47} and implemented using the snompy package \cite{ref49}. For a direct comparison, we use the same geometrical parameters, R, L, A, and H0, as in the MESM, while the phenomenological complex-valued \(g\)-coefficient is fitted to the experimental nano-FTIR spectra. We found that for weakly resonant samples such as PMMA, the FDM is just as insensitive to spheroid length as the ESM. Furthermore, because the FDM represents the tip as a finite set of charges, the spectrum calculated using this model depends very weakly on \(H_{0}\), in contrast to the ESM. As a result, within the FDM, small variations in g can compensate for variations in \(H_{0}\) and \(L\). The best agreement is obtained for \(g = 0.76e^{0.07i}\). With this value, the FDM can reproduce a selected PMMA spectrum reasonably well. However, when the same \(g\) and geometrical parameters are subsequently used for the other layer thicknesses and harmonics, substantial discrepancies emerge (Figures 2e,f). In particular, no single value of \(g\) provides a simultaneous quantitative description of the complete experimental dataset. This behavior is consistent with the limitations identified for the FDM already in the bulk-sample case \cite{ref50}, where a parameter set optimized for one harmonic does not necessarily reproduce other harmonics with the same accuracy. For multilayer samples, the additional dependence on layer thickness makes this limitation substantially more pronounced. Figure 2g summarizes the comparison by showing the modulus of the difference between the calculated and measured fourth-harmonic near-field amplitudes (top) and phases (bottom), defined as \(\left| \mathrm{\Delta}s_{4} \right| = \left| s_{4}^{calc} - s_{4}^{\exp} \right|\) and \(\left| \mathrm{\Delta}\varphi_{4} \right| = \left| \varphi_{4}^{calc} - \varphi_{4}^{\exp} \right|\), respectively. Across the investigated PMMA thicknesses, the differences between the calculated and measured spectra are up to one order of magnitude larger for the FDM than for the MESM.

The ability to reproduce spectra measured at different layer thicknesses and demodulation harmonics using a single, physically consistent set of model parameters is essential for quantitative s-SNOM spectroscopy to reconstruct the sample\textquotesingle s optical properties, particularly for tomography \cite{ref33,ref34,ref35}. In contrast to fitting individual spectra independently, such parameter transferability allows variations in the measured response to be attributed to the intrinsic optical properties and geometry of the sample rather than to readjustment of phenomenological model parameters. Combined with its low computational cost, the MESM therefore provides a practical framework for systematic parameter optimization and inverse modeling of layered samples, including the extraction of local dielectric functions and other material parameters from experimental near-field spectra.

To test the model for strongly resonant polaritonic systems, we used nano-FTIR measurements on a 50-nm-thick SiC layer on an InAs substrate reported in ref. \cite{ref58}. The geometry of the model is illustrated in Figure 3a. SiC is a polar dielectric exhibiting a strong mid-infrared optical response associated with its optical phonons; within the Reststrahlen band, its dielectric permittivity becomes negative (Figure 3b) \cite{ref55,ref58}. This makes modeling the SiC layer considerably more challenging than modeling a weak molecular resonance. The InAs substrate provides an additional means of testing the consistency of the model. Without photoexcitation, InAs behaves as a dielectric in the investigated mid-infrared spectral range, with a real positive dielectric permittivity (Figure 3c, top). Upon excitation by a near-infrared pump pulse, photoinduced free carriers introduce a pronounced Drude response, transiently switching the substrate to a metallic state in the same spectral range (Figure 3a, right, and Figure 3c, bottom), dramatically changing the optical response of the SiC/InAs heterostructure. Importantly, the geometrical parameters of the tip--sample system---including the tip radius, spheroid length, tapping amplitude, and effective minimum tip--sample separation---remain unchanged between the non-excited and photoexcited measurements, providing a particularly stringent test of the model parameters\textquotesingle{} transferability. In all calculations, we use the dielectric constant from \cite{ref58}, which is also provided in Section S10 of the Supporting Information.

We first consider SiC on non-excited InAs. Figure 3d compares the measured amplitude and phase spectra of the second- and third-harmonic near-field signals with the corresponding MESM calculations, while Figure 3e shows the same comparison for the FDM. For the MESM, we use \(R = 75\,\mathrm{nm}\), a reasonable value for a nano-FTIR tip used in the measurements, and \(A = 75\,\mathrm{nm}\), a nominal value set in the experiments. As is known, for highly resonant materials, spheroid length significantly affects the spectrum \cite{ref37,ref50}. We choose \(L = 2.0\,\mu\mathrm{m}\) to best fit the obtained spectra to the experimental data. Notably, the submicron lengths typically used in simulations yielded a poorer fit than the chosen higher value, which is closer to the actual tip dimensions. However, further increasing the spheroid length is incorrect within this model, as it makes \(L\) close to the wavelength and violates the applicability conditions of the electrostatic approximation. All spectra are normalized to the near-field response of bulk, non-excited InAs, which is calculated using \(H_{0\mathrm{InAs}} = 0.1\,\mathrm{nm}\) and the same tip geometrical parameters as for the SiC/InAs structure. In contrast to the weakly resonant PMMA system considered above, however, calculation of the SiC response requires additional treatment of narrow spectral features arising in the strongly resonant tip--sample interaction, as described in the following paragraph.

For a fixed minimum tip--sample distance \(H_{0}\), the calculated SiC spectra exhibit several narrow resonances associated with strongly confined gap modes formed between the tip apex and the sample \cite{ref37,ref50,ref59}. Their spectral positions and amplitudes depend sensitively on \(H_{0}\). These resonances and the corresponding gap-mode field distributions are illustrated explicitly in Section S7 of the Supporting Information. Such fine spectral structure is generally not resolved in experimental s-SNOM spectra \cite{ref8,ref60,ref61}. We therefore account phenomenologically for small variations of the effective tip--sample separation, arising from tip-motion fluctuations, nanoscale roughness, and deviations of the actual tip apex from the ideal spheroidal geometry, by averaging the calculated spectra over \(H_{0\mathrm{SiC}} = 1 - 5\,\mathrm{nm}\). The averaged spectra are subsequently convolved with a Gaussian function with a FWHM of 12.5 cm\textsuperscript{-1}, corresponding to the interferometer bandwidth, following the procedure used for strongly resonant quartz in our previous work \cite{ref50}. This treatment is not required for the substantially smoother PMMA spectra. Owing to the computational efficiency of the MESM, the complete \(H_{0}\)-averaged spectrum can still be calculated within several seconds.

With this procedure, the MESM reproduces well the experimental spectra for both measured signal harmonics (Figure 3d). In particular, it captures well the characteristic two peaks within the Reststrahlen band, which correspond to lower and upper surface phonon polariton resonance of the SiC layer \cite{ref58}. For comparison, we perform the FDM calculations shown in Figure 3e using the same curvature radius and tapping amplitude as the MESM: \(R = 75\,\mathrm{nm}\), \(A = 75\,\mathrm{nm}\). For the minimal tip-sample distance, we do not perform the averaging procedure because the FDM spectra do not significantly depend on \(H_{0}\) varying from 1 to 5 nm, as this model does not capture tip-sample gap modes. Therefore, we choose a single constant value: \(H_{0\mathrm{SiC}} = H_{0\mathrm{InAs}} = 1\,\mathrm{nm}\). Using \(L\) and \(g\) as fitting parameters, we obtain the best match to the measured spectra with \(g = 1.6e^{0.27i}\) and \(L = 400\,\mathrm{nm}\). Even after optimization of \(g\) and \(L\) for non-excited InAs, the FDM does not reproduce the lower-frequency peak with the same accuracy and exhibits rapid spectral oscillations in parts of the calculated response, indicating poor convergence in this strongly resonant regime. Thus, unlike in the weak-oscillator case, the limitations of the phenomenological approach affect not only the quantitative magnitude of the near-field signal but also its spectral line shape. Note that the obtained spheroid length is significantly smaller than the one used for the MESM, whereas using larger \(L\) leads to additional peaks absent in the experimental spectra. This discrepancy suggests that although the FDM is based on spheroid geometry, it is an approximate solution, which can reproduce the exact solution with different accuracy depending on the parameter values. Because longer spheroids distribute charge more widely, the discrete-charge approximation used in the FDM becomes less applicable.

\begin{figure}[!t]
\centering
\includegraphics[width=\linewidth,height=0.49\textheight,keepaspectratio]{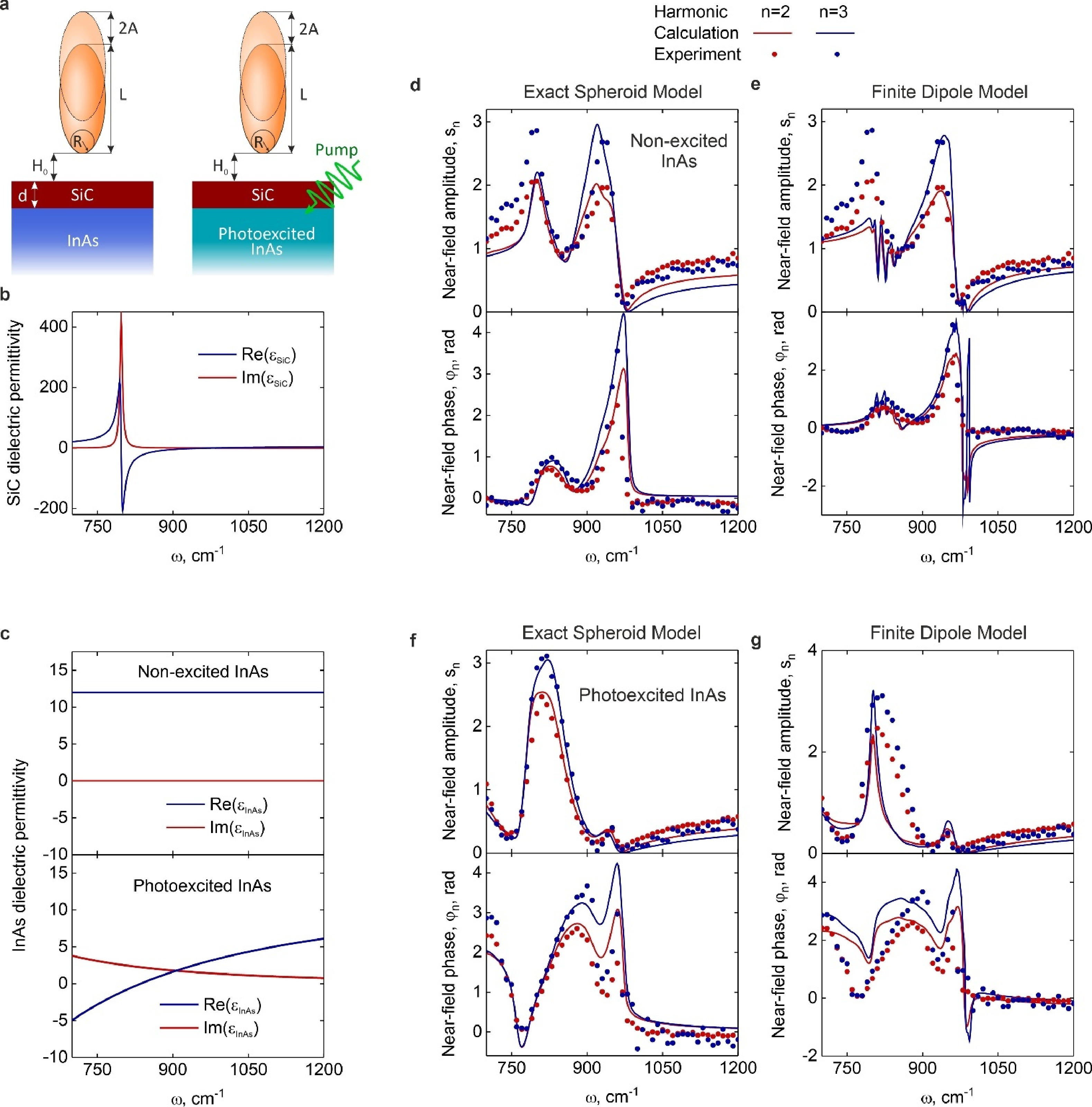}
\caption{a) Schematic of the prolate spheroid above a 50-nm-thick SiC layer on an InAs substrate. In the absence of photoexcitation, InAs exhibits a dielectric response (left), whereas excitation by a near-infrared pump pulse generates free carriers and transiently switches the substrate to a metallic state (right). The geometrical parameters of the model are \(R = 75\,\mathrm{nm}\), \(L = 2.0\,\mu\mathrm{m}\), and \(A = 75\,\mathrm{nm}\). To account phenomenologically for fluctuations of the effective tip--sample separation and nanoscale roughness, the minimum distance between the tip and the SiC surface, \(H_{0\mathrm{SiC}}\), is varied from 1 to 5 nm and the resulting spectra are averaged. All spectra are normalized to the near-field response of bulk, non-excited InAs, \(H_{0\mathrm{InAs}} = 0.1\,\mathrm{nm}\). b) Dielectric permittivity of SiC as a function of incident-light frequency, taken from \cite{ref58}. c) Dielectric permittivity of non-excited (top) and photoexcited (bottom) InAs, taken from \cite{ref58}. d,e) Amplitude and phase of the second- and third-harmonic near-field spectra of SiC on non-excited InAs, calculated using the MESM (d) and the FDM (e), respectively. f,g) Corresponding spectra for photoexcited InAs, calculated using the MESM (f) and the FDM (g), respectively. Solid curves represent the calculated near-field signal, and symbols represent the experimental data. The MESM calculations use the same geometrical parameters and \(H_{0}\)-averaging procedure for both non-excited and photoexcited InAs, whereas the FDM uses the same \(R = 75\,\mathrm{nm}\), \(L = 400\,\mathrm{nm}\), and \(A = 75\,\mathrm{nm}\), but requires independently fitted \(g\)-coefficients for the two substrate states: \(g = 1.6e^{0.27i}\) (e) and \(g = 1.65e^{0.007i}\) (g).}
\label{fig:3}
\end{figure}

We next consider the same SiC/InAs structure after photoexcitation of the InAs substrate. Photoexcitation strongly modifies the measured near-field response due to ultrastrong coupling between tip-excited surface plasmons in photoexcited InAs and the TO phonon in SiC (Figure 3f) \cite{ref58}. The MESM reproduces this modified spectrum using exactly the same geometrical parameters and the same \(H_{0}\)-averaging procedure as for non-excited InAs. No refitting of the tip geometry is required. This is particularly important because the two measurements are performed on the same physical structure and differ only in the carrier density of the InAs substrate induced by photoexcitation. The situation is qualitatively different for the FDM. If the FDM parameters optimized for non-excited InAs are used to calculate the spectrum of SiC on photoexcited InAs, the calculated near-field signal differs qualitatively from the experimental signal and does not reproduce the main resonance peak, as shown in Section S10 of the Supporting Information. To reproduce the experimental spectrum, the FDM \(g\)-coefficient has to be refitted. The best agreement is obtained for \(g = 1.65e^{0.007i}\); for \(R\), \(A\), and \(L\), we use the same values as for non-excited InAs. After this adjustment, the FDM provides the qualitatively correct spectral shape and quantitatively fits the main-peak magnitude, as shown in Figure 3g. However, the FDM spectrum still matches the experimental one significantly worse than the MESM result. Therefore, although independently fitted g-parameters can provide a good qualitative agreement with experimental spectra for each state of the InAs substrate, the FDM does not have a single transferable set of parameters for both states of the same SiC/InAs system. In contrast, the MESM describes both optical states using one common set of geometrical parameters.

The results presented above demonstrate the substantial advantage of the MESM for quantitative s-SNOM spectroscopy. In particular, they show that the same model parameters can be used when the sample properties are changed, without refitting the model. In the PMMA measurements, the layer thickness is varied, while in the SiC/InAs experiment, the optical response of the InAs substrate is changed by photoexcitation while the physical structure remains unchanged. In both cases, the MESM describes the different measured spectra using the same model parameters. In contrast, the FDM requires refitting its g-coefficient when the sample response is changed, as demonstrated for the photoexcited InAs substrate. Thus, the FDM does not provide a single set of parameters that describes both measurements. The ability of the MESM to describe different sample responses without changing its parameters is of critical importance for quantitative spectroscopy and, in particular, for inverse problems, where the optical properties and layer thickness of a sample are determined from a set of near-field spectra measured at higher harmonics \cite{ref33,ref34,ref35,ref62}.

\section{Conclusion and outlook}

We have generalized the exact quasi-electrostatic spheroid model of s-SNOM \cite{ref50} to planar multilayer samples by introducing the GICM. The GICM represents the wavevector-dependent electrostatic response of a layered structure by a systematically convergent series of image charges and can be incorporated directly into the ESM developed previously for homogeneous samples. The resulting MESM retains the computational efficiency of an analytical approach while reproducing FEM calculations within numerical precision. We experimentally validated the model in two complementary regimes. For weakly resonant PMMA films, a single set of geometrical parameters simultaneously reproduces the measured spectra for different film thicknesses and demodulation harmonics. For strongly resonant SiC/InAs structures, the model describes both non-excited and photoexcited states of the InAs substrate using the same geometrical parameters, despite the pronounced change in the optical response of the sample. In contrast, the FDM requires readjustment of its \(g\)-parameter and does not provide a quantitatively consistent description of the complete datasets. These results demonstrate that the MESM provides an efficient framework for quantitative analysis of s-SNOM spectra of layered samples and creates a basis for parameter optimization and inverse reconstruction of their local optical properties.

Several limitations and open questions nevertheless remain. As in other spheroidal-tip models, the effective spheroid length \(L\) is the only geometrical parameter that cannot be directly identified with a well-defined parameter of the real s-SNOM tip, and establishing its relation to the actual tip geometry remains an important problem that requires additional experimental and theoretical research. The minimum tip--sample distance \(H_{0}\) represents another major source of uncertainty because it cannot generally be determined with nanometer precision in conventional s-SNOM setups. Approaches combining near-field optical detection with scanning tunneling microscopy (STM) may provide substantially better control of this parameter \cite{ref15,ref16}. Furthermore, in the present work, fluctuations of the effective tip--sample distance and nanoscale tip and sample roughness are treated phenomenologically by averaging spectra over a range of \(H_{0}\). Whether this procedure provides a sufficiently accurate description of the different sources of experimental noise and geometrical imperfections requires further investigation. These uncertainties are particularly relevant because the near-field response can depend strongly on \(H_{0}\), as already established for the bulk spheroid model. We emphasize that these parameters play a special role in modeling near-field spectroscopy of highly resonant, low-loss polariton samples, such as SiC, whose spectra are influenced by tip-sample gap modes. Fitting the spectra of such materials requires special attention when selecting model parameters.

Beyond the approximation of the actual tip by a prolate spheroid, the present theory relies on the quasi-electrostatic approximation. While this approximation is well justified for the mid-infrared conditions investigated here, retardation effects are expected to become increasingly important toward the near-infrared and visible spectral ranges, where a fully electrodynamic treatment or perturbative electrodynamic corrections may be required. Further extensions of the model could also include multilayers with in-plane optical anisotropy and nonlocal optical responses of either the sample or the tip. Addressing these effects, together with improved experimental determination of the relevant geometrical parameters, should further increase the predictive capability of quantitative s-SNOM modeling.

To facilitate the application and further development of the model, we provide the implementation as open-source code. In addition, we provide a user-friendly interface that enables rapid calculation and visualization of s-SNOM spectra for commonly encountered sample structures; further details are provided in Section S11 of the Supporting Information \cite{ref63}. This combination of computational efficiency, quantitative accuracy, and accessibility should facilitate the use of the model for forward spectral calculations and, ultimately, for extracting local material properties from experimental near-field measurements.

\section*{Acknowledgements}
We thank Martin Schnell for fruitful and constructive discussions. We thank Petr Andreev for technical assistance during the early-stage development of a web interface for the s-SNOM spectrum calculator. The authors used OpenAI ChatGPT 5.6 and 6 Sol as text-editing tools to improve grammar and sentence structure, enhancing the text\textquotesingle s clarity and readability. Claude Code with Claude Opus 5.5 and Sonnet 5.5 served as a coding companion to improve efficiency during script optimization, FIM implementation, and user interface development. The authors fully reviewed all AI-processed content, manually tested and verified the final code, and remained fully accountable for the content. E.V. acknowledges funding from the European Union\textquotesingle s Horizon 2020 research and innovation program under the Marie Sk\l{}odowska-Curie grant agreement No. 101063152. E.V. and R.H. acknowledge support from grant PID2024-156602NB-I00 (CEDANO) and grant CEX2020-001038-M funded by MICIU/ AEI/10.13039/501100011033 and by ERDF/EU. A.Y.N. acknowledges support from the Spanish Ministry of Science and Innovation through Grant PID2023-147676NBI00, the Basque Department of Education through Grant PIBA-2023-1-0007, and the IKUR Strategy.

\end{document}


\title{Supporting Information for "Quantitative Analytical Spheroid Model for Scattering-Type Scanning Near-Field Optical Spectroscopy of Layered Samples"}

\newcommand{\dipc}{
Donostia International Physics Center (DIPC), Donostia/San-Sebastian, Spain
}

\newcommand{\NanoGUNE}{
CIC nanoGUNE BRTA, Tolosa Hiribidea 76, 20018 Donostia-San Sebastian, Spain
}

\newcommand{\Attocube}{
Attocube Systems GmbH, 85540 Haar, Germany
}

\newcommand{\IKERBASQUE}{
IKERBASQUE, Basque Foundation for Science, Bilbao 48013, Spain
}

\author{Kirill V. Voronin}
\email[]{kirill.voronin@dipc.org}
\affiliation{\dipc}

\author{Edoardo Vicentini}
\affiliation{\NanoGUNE}

\author{Lars Mester}
\affiliation{\Attocube}

\author{Rainer Hillenbrand}
\email[]{r.hillenbrand@nanogune.eu}
\affiliation{\NanoGUNE}
\affiliation{\IKERBASQUE}

\author{Alexey Y. Nikitin}
\email[]{alexey@dipc.org}
\affiliation{\dipc}
\affiliation{\IKERBASQUE}

\maketitle

{
  \hypersetup{linkcolor=black}
  \tableofcontents
}

\newpage

\section{Exact spheroid model solution for a bulk sample}

Here, we briefly summarize the main steps in deriving the polarizability of a prolate spheroid placed above a homogeneous bulk sample in an external electric field \cite{ESM}. This formulation provides the starting point for the generalization to multilayer samples developed in the following sections. In addition to approximating the s-SNOM tip by a prolate spheroid, we employ the quasi-electrostatic approximation, in which retardation and magnetic-field effects are neglected and the electromagnetic problem reduces to an electrostatic boundary-value problem. Within this approximation, the incident field can be treated as spatially uniform over the spheroid. We consider excitation polarized along the spheroid major axis, since the elongated tip geometry strongly enhances the response to the vertical electric-field component.

\begin{figure}[H]
\centering
\includegraphics[width=12cm]{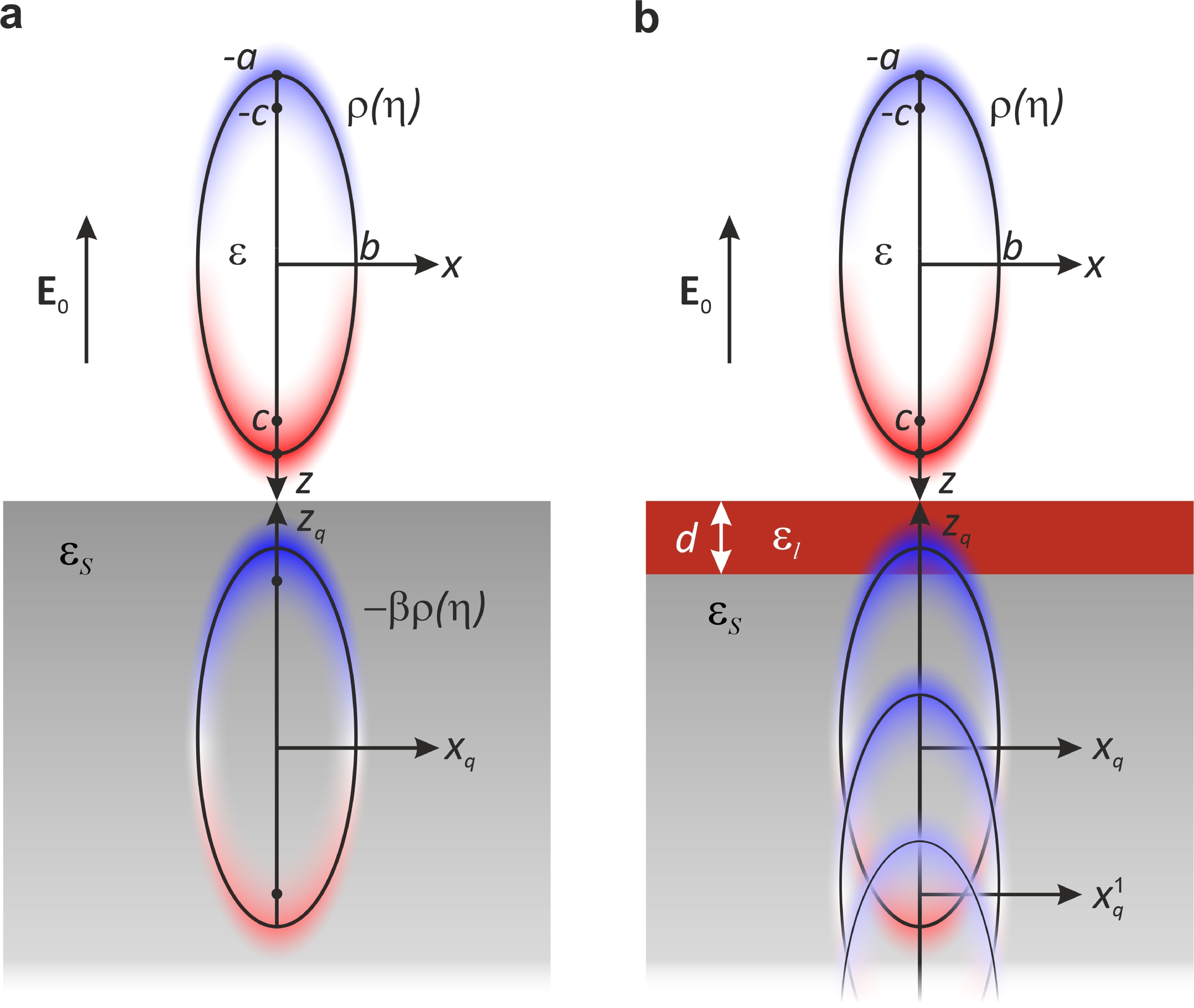}
\caption{a) Schematic of a spheroid with dielectric permittivity $\eps$ placed above a homogeneous bulk sample with dielectric permittivity $\eps_s$ in a uniform vertical electric field $\mathbf{E}_0$. Here, $a$ and $b$ are the major and minor semi-axes of the spheroid, respectively, while $c$ and $-c$ denote the positions of its focal points. The surface charge density is denoted by $\rho(\eta)$. Within the electrostatic image method, the response of the bulk sample can be represented by an image spheroid with surface charge density $-\beta\rho(\eta)$. b) Same geometry for a spheroid placed above a layer of thickness $d$ and dielectric permittivity $\eps_l$ on a substrate with dielectric permittivity $\eps_s$. In this case, the electrostatic response of the layered sample is represented by a series of spheroid images.}
\label{Ellipsoid_scheme}
\end{figure}

To solve the electrostatic problem, it is convenient to introduce prolate spheroidal coordinates, $(\mu,\nu,\phi)$, associated with the spheroid shown in Figure \ref{Ellipsoid_scheme}a:
\begin{equation}
\begin{array}{l}
x=c \sinh{\mu} \sin{\nu} \cos{\phi}, \\
y=c \sinh{\mu} \sin{\nu} \sin{\phi}, \\
z=c \cosh{\mu} \cos{\nu},
\end{array}
\label{Sph_coord}
\end{equation}
where $c$ is the distance from the spheroid center to either focal point, and the coordinates span $\mu \in [0,\infty)$, $\nu \in [0,\pi]$, and $\phi \in [0,2\pi]$. The spheroid surface corresponds to $\mu=\mu_0$. Its major semi-axis $a=L/2$ and minor semi-axis $b$ are related to $\mu_0$ through
$a=c\cosh{\mu_0}$ and $b=c\sinh{\mu_0}$, respectively. The radius of curvature at the spheroid apex is therefore
\begin{equation}
R=\frac{b^2}{a}
=c\sinh{\mu_0}\tanh{\mu_0}.
\label{Sph_rad}
\end{equation}

For convenience, we introduce
\begin{equation}
\xi=\cosh{\mu}, \qquad \eta=\cos{\nu}.
\end{equation}
Because both the geometry and the excitation are axially symmetric, the electrostatic potential is independent of $\phi$. Solutions of Laplace's equation can therefore be expanded in Legendre functions of the first and second kind, $P_n(x)$ and $Q_n(x)$, respectively. The potential generated by the polarization charges on the spheroid and evaluated outside it must decay at infinity and can be written as
\begin{equation}
\varphi_r(\xi,\eta)
=
\sum_{n=0}^\infty A_n Q_n(\xi)P_n(\eta),
\label{Sph_Laplace_Sol_Out}
\end{equation}
whereas the potential inside the spheroid, which must remain finite, can be expanded as
\begin{equation}
\varphi_{in}(\xi,\eta)
=
\sum_{n=0}^\infty B_n P_n(\xi)P_n(\eta).
\label{Sph_Laplace_Sol_In}
\end{equation}

The total potential outside the spheroid is
\begin{equation}
\varphi_{out}=\varphi_e+\varphi_r,
\end{equation}
where $\varphi_r$ is the potential generated by the polarization charges on the spheroid and $\varphi_e$ is the external potential acting on it. The latter contains the potential of the incident field,
\begin{equation}
\varphi_0=cE_0\xi\eta,
\end{equation}
where $E_0$ is the incident field amplitude, and the potential $\varphi_s$ produced by the polarization charges induced in the sample, so that $\varphi_e=\varphi_0+\varphi_s$. The electrostatic boundary conditions at the spheroid surface, $\xi=\xi_0$, are therefore
\begin{equation}
\varphi_{\text{in}}(\xi_0,\eta)
=
\varphi_{\text{out}}(\xi_0,\eta)
=
\varphi_e(\xi_0,\eta)+\varphi_r(\xi_0,\eta),
\label{Sph_BC_0}
\end{equation}
and
\begin{equation}
\eps
\left.
\frac{\partial\varphi_{\text{in}}(\xi,\eta)}{\partial\xi}
\right|_{\xi_0}
=
\left.
\frac{\partial\varphi_{\text{out}}(\xi,\eta)}{\partial\xi}
\right|_{\xi_0},
\label{Sph_BC_1}
\end{equation}
where $\eps$ is the dielectric permittivity of the spheroid. Equation \eqref{Sph_BC_0} expresses continuity of the electrostatic potential, whereas Equation \eqref{Sph_BC_1} follows from continuity of the normal component of the electric displacement field, $\bD=-\eps\nabla\varphi$.

Owing to the axial symmetry of the problem, the surface charge density on the spheroid depends only on $\eta$ and can be expanded in Legendre polynomials as
\begin{equation}
\rho(\eta)
=
\frac{
\sum\limits_{n=0}^\infty
(2n+1)\upsilon_n P_n(\eta)
}{
2\pi c
\sqrt{(\xi_0^2-1)(\xi_0^2-\eta^2)}
}.
\label{Sph_sig}
\end{equation}
The electrostatic potential generated by this surface charge distribution outside the spheroid is then
\begin{equation}
\varphi_r(\xi,\eta)
=
2
\sum\limits_{n=0}^\infty
(2n+1)\upsilon_n
P_n(\xi_0)P_n(\eta)Q_n(\xi).
\label{Sph_phi_alpha}
\end{equation}

It is useful to separate the total surface charge into two contributions,
\begin{equation}
\rho(\eta)=\rho_0(\eta)+\rho_i(\eta),
\end{equation}
where $\rho_0$ is the charge induced by the uniform external field in the absence of the sample, and $\rho_i$ is the additional charge induced by the electrostatic interaction with the sample. The former can be obtained analytically:
\begin{equation}
\rho_0(\eta)
=
\frac{
3\upsilon^{(0)}_1P_1(\eta)
}{
2\pi c
\sqrt{(\xi_0^2-1)(\xi_0^2-\eta^2)}
}
=
-\frac{
(\eps-1)E_0\eta
}{
4\pi
\left(
\eps Q_1(\xi_0)
-\xi_0Q_0(\xi_0)
+\frac{\xi_0^2}{\xi_0^2-1}
\right)
\sqrt{(\xi_0^2-1)(\xi_0^2-\eta^2)}
}.
\label{Sph_sig_0_eps}
\end{equation}

The remaining task is therefore to determine $\rho_i(\eta)$. For a homogeneous semi-infinite sample, the electrostatic interaction can be treated exactly using the image method \cite{jackson1999classical}. For a sample with dielectric permittivity $\eps_s$, we define
\begin{equation}
\beta=\frac{\eps_s-1}{\eps_s+1}.
\end{equation}
Each polarization charge on the spheroid then produces an image charge below the sample surface with amplitude $-\beta$ times that of the original charge. Consequently, the sample-induced potential is equivalent to the potential generated by an image spheroid carrying the surface charge density $-\beta\rho(\eta)$.

Using the spheroidal-harmonic expansion of the Coulomb potential, the sample-induced potential can be written as
\begin{multline}
\varphi_s(\xi,\eta,\phi)
=
-\beta
\int\limits_{-1}^1
c^2
\sqrt{(\xi_0^2-1)(\xi_0^2-\eta^{\prime 2})}
\,d\eta^{\prime}
\int\limits_0^{2\pi}
d\phi^{\prime}
\,
\frac{\rho_i(\eta^{\prime})+\rho_0(\eta^{\prime})}{c}
\\
\times
\sum\limits_{n=0}^\infty
(2n+1)
\sum\limits_{m=0}^n
\epsilon_m i^m
\left[
\frac{(n-m)!}{(n+m)!}
\right]
\cos{m(\phi+\phi^{\prime})}
P_n^m(\eta)
P_n^m(\eta_q)
P_n^m(\xi)
Q_n^m(\xi_q),
\label{Sph_phi_i_sigma}
\end{multline}
where $\epsilon_m=2-\delta_{m0}$ is the Neumann factor, while $\eta_q$ and $\xi_q$ denote the spheroidal coordinates of the corresponding image point expressed in the coordinate system associated with the real spheroid.

The relation between $(\xi_q,\eta_q)$ and the coordinates $(\xi^{\prime},\eta^{\prime})$ of the original surface charge, with $\xi^{\prime}=\xi_0$, can be obtained in Cartesian coordinates, as illustrated in Figure \ref{Ellipsoid_scheme}. For the $z$ coordinate,
\begin{equation}
c\xi_q\eta_q
=
z
=
2a+2H-z^{\prime}
=
2a+2H-c\xi_0\eta^{\prime},
\label{Sph_z}
\end{equation}
whereas for the transverse coordinate,
\begin{equation}
c\sqrt{\xi_q^2-1}\sqrt{1-\eta_q^2}
=
x
=
c\sqrt{\xi_0^2-1}\sqrt{1-\eta^{\prime 2}}.
\label{Sph_x}
\end{equation}
Solving these relations gives
\begin{equation}
\eta_q=\frac{\psi}{\xi_q},
\qquad
\xi_q
=
\sqrt{
\frac{
1+\psi^2+\chi^2
+
\sqrt{(1+\psi^2+\chi^2)^2-4\psi^2}
}{2}
},
\label{Sph_eta_xi_i}
\end{equation}
where
\begin{equation}
\psi
=
2\frac{a+H}{c}
-\xi_0\eta^{\prime},
\qquad
\chi
=
\sqrt{(\xi_0^2-1)(1-\eta^{\prime 2})}.
\label{Sph_psi_chi}
\end{equation}

We now expand the interaction-induced charge $\rho_i$ in the form of Equation \eqref{Sph_sig}, with expansion coefficients $\upsilon_n$. Substituting this expansion together with $\rho_0$ from Equation \eqref{Sph_sig_0_eps} into Equation \eqref{Sph_phi_i_sigma}, and carrying out the integration over $\phi^{\prime}$, yields the sample-induced potential at the spheroid surface:
\begin{multline}
\varphi_s(\xi_0,\eta)
=
-\beta
\sum\limits_{n=0}^\infty
\sum\limits_{m=0}^\infty
(2n+1)(2m+1)
\upsilon_n
J_{nm}
P_m(\eta)P_m(\xi_0)
\\
+
\beta
\frac{
c(\eps-1)E_0
}{
2
\left(
\eps Q_1(\xi_0)
-\xi_0Q_0(\xi_0)
+\frac{\xi_0^2}{\xi_0^2-1}
\right)
}
\sum\limits_{m=0}^\infty
(2m+1)
J_{1m}
P_m(\eta)P_m(\xi_0),
\label{Sph_phi_i_sigma_2_eps}
\end{multline}
where we have introduced the geometrical coupling integral
\begin{equation}
J_{nm}
=
\int\limits_{-1}^1
d\eta\,
P_n(\eta)
P_m(\eta_q)
Q_m(\xi_q).
\label{Sph_int_J_sig}
\end{equation}
The quantity $J_{nm}$ describes the electrostatic coupling between the $n^{\texttt{th}}$-order multipole component on the real spheroid and the $m^{\texttt{th}}$-order component generated by its image. Importantly, $J_{nm}$ depends only on the system geometry and can therefore be calculated independently of the sample's optical properties.

Since the free-spheroid solution $(\varphi_0,\rho_0)$ already satisfies the boundary conditions in the absence of the sample, it can be subtracted from the full boundary-value problem. The remaining equations then determine only the sample-induced correction $\rho_i$. Using the representation of the potential inside the spheroid given by Equation \eqref{Sph_Laplace_Sol_In}, the continuity condition in Equation \eqref{Sph_BC_0} becomes
\begin{equation}
2
\sum\limits_{m=0}^\infty
(2m+1)
\upsilon_m
P_m(\xi_0)P_m(\eta)Q_m(\xi_0)
+
\varphi_s(\xi_0,\eta)
=
\sum\limits_{m=0}^\infty
B_mP_m(\xi_0)P_m(\eta).
\label{Sph_BC_for_sigma_i_eps}
\end{equation}
Using the orthogonality of the Legendre polynomials, this gives
\begin{equation}
2\upsilon_m Q_m(\xi_0)
-
\beta
\sum\limits_{n=0}^\infty
(2n+1)\upsilon_nJ_{nm}
+
\beta
\frac{
c(\eps-1)E_0J_{1m}
}{
2
\left(
\eps Q_1(\xi_0)
-\xi_0Q_0(\xi_0)
+\frac{\xi_0^2}{\xi_0^2-1}
\right)
}
=
\frac{B_m}{2m+1}.
\label{Sph_Eq_up_n1_eps}
\end{equation}

Applying the second boundary condition, Equation \eqref{Sph_BC_1}, similarly gives
\begin{multline}
2\upsilon_mP_m(\xi_0)
\frac{
\xi_0Q_m(\xi_0)-Q_{m-1}(\xi_0)
}{
\xi_0P_m(\xi_0)-P_{m-1}(\xi_0)
}
-
\beta
\sum\limits_{n=1}^\infty
(2n+1)\upsilon_nJ_{nm}
\\
+
\beta
\frac{
c(\eps-1)E_0J_{1m}
}{
2
\left(
\eps Q_1(\xi_0)
-\xi_0Q_0(\xi_0)
+\frac{\xi_0^2}{\xi_0^2-1}
\right)
}
=
\frac{\eps B_m}{2m+1}.
\label{Sph_Eq_BC2_sig_eps}
\end{multline}
Eliminating $B_m$ using Equation \eqref{Sph_Eq_up_n1_eps} yields
\begin{multline}
2\upsilon_m
\left[
\eps\frac{Q_m(\xi_0)}{P_m(\xi_0)}
-
\frac{
\xi_0Q_m(\xi_0)-Q_{m-1}(\xi_0)
}{
\xi_0P_m(\xi_0)-P_{m-1}(\xi_0)
}
\right]
-
\beta(\eps-1)
\sum\limits_{n=1}^\infty
(2n+1)\upsilon_n
\frac{J_{nm}}{P_m(\xi_0)}
= \\ =
-\beta
\frac{
c(\eps-1)^2E_0J_{1m}
}{
2P_m(\xi_0)
\left(
\eps Q_1(\xi_0)
-\xi_0Q_0(\xi_0)
+\frac{\xi_0^2}{\xi_0^2-1}
\right)
}.
\label{Sph_eps_Eq_sig}
\end{multline}

For a neutral spheroid, the monopole coefficient vanishes, $\upsilon_0=0$, and the remaining unknown coefficients correspond to $n\geq1$. Dividing Equation \eqref{Sph_eps_Eq_sig} by $E_0$ and introducing
\begin{equation}
\vartheta_n=\frac{\upsilon_n}{E_0},
\end{equation}
the problem can be written as the linear system
\begin{equation}
\sum\limits_{n=1}^{\infty}
\mathcal{M}_{mn}\vartheta_n
=
\mathcal{C}_m,
\qquad m\geq1,
\label{Sph_syst_eps_sig}
\end{equation}
where
\begin{equation}
\mathcal{M}_{mn}
=
\delta_{nm}
-
\frac{
\beta(\eps-1)(2n+1)J_{nm}
}{
2P_m(\xi_0)
\left[
\eps\frac{Q_m(\xi_0)}{P_m(\xi_0)}
-
\frac{
\xi_0Q_m(\xi_0)-Q_{m-1}(\xi_0)
}{
\xi_0P_m(\xi_0)-P_{m-1}(\xi_0)
}
\right]
},
\label{Sph_Mat_eps_sig}
\end{equation}
and
\begin{equation}
\mathcal{C}_m
=
-
\frac{
\beta c(\eps-1)^2J_{1m}
}{
4P_m(\xi_0)
\left(
\eps Q_1(\xi_0)
-\xi_0Q_0(\xi_0)
+\frac{\xi_0^2}{\xi_0^2-1}
\right)
\left[
\eps\frac{Q_m(\xi_0)}{P_m(\xi_0)}
-
\frac{
\xi_0Q_m(\xi_0)-Q_{m-1}(\xi_0)
}{
\xi_0P_m(\xi_0)-P_{m-1}(\xi_0)
}
\right]
}.
\label{Sph_C_eps_sig}
\end{equation}
In practical calculations, the infinite system in Equation \eqref{Sph_syst_eps_sig} is truncated at a sufficiently large multipole order, which is chosen by verifying convergence of the calculated polarizability.

The final step is to calculate the dipole moment of the prolate spheroid integrating $z\rho$ over its surface $S$. Using $z=c\xi_0\eta$ and the surface element
\begin{equation}
dS
=
c^2
\sqrt{(\xi_0^2-1)(\xi_0^2-\eta^2)}
\,d\eta\,d\phi,
\end{equation}
the vertical dipole moment is 
\begin{equation}
p_z = \int\limits_S z \rho dS =  2\pi c^3\xi_0
\int\limits_{-1}^1
\eta\,\rho(\eta)
\sqrt{(\xi_0^2-1)(\xi_0^2-\eta^2)}
\,d\eta.
\label{PZ}
\end{equation}
Substituting $\rho=\rho_i+\rho_0$, using Equations \eqref{Sph_sig} and \eqref{Sph_sig_0_eps}, and dividing the resulting dipole moment by $E_0$, gives the vertical polarizability of the spheroid:
\begin{equation}
\alpha_z
=
2ac\vartheta_1
-
\frac{
(\eps-1)c^3P_1\left(\frac{a}{c}\right)
}{
3
\left[
\eps Q_1\left(\frac{a}{c}\right)
-
P_1\left(\frac{a}{c}\right)
Q_0\left(\frac{a}{c}\right)
+
\frac{a^2}{a^2-c^2}
\right]
}.
\label{Sph_alphaZ_eps_sig}
\end{equation}
Together with the solution of Equation \eqref{Sph_syst_eps_sig}, this expression gives the exact quasi-electrostatic polarizability of a prolate spheroid above a homogeneous bulk sample and provides the basis for the multilayer generalization developed in the following sections.

\section{Electrostatic image method}

\begin{figure}[H]
\centering
\includegraphics[width=16cm]{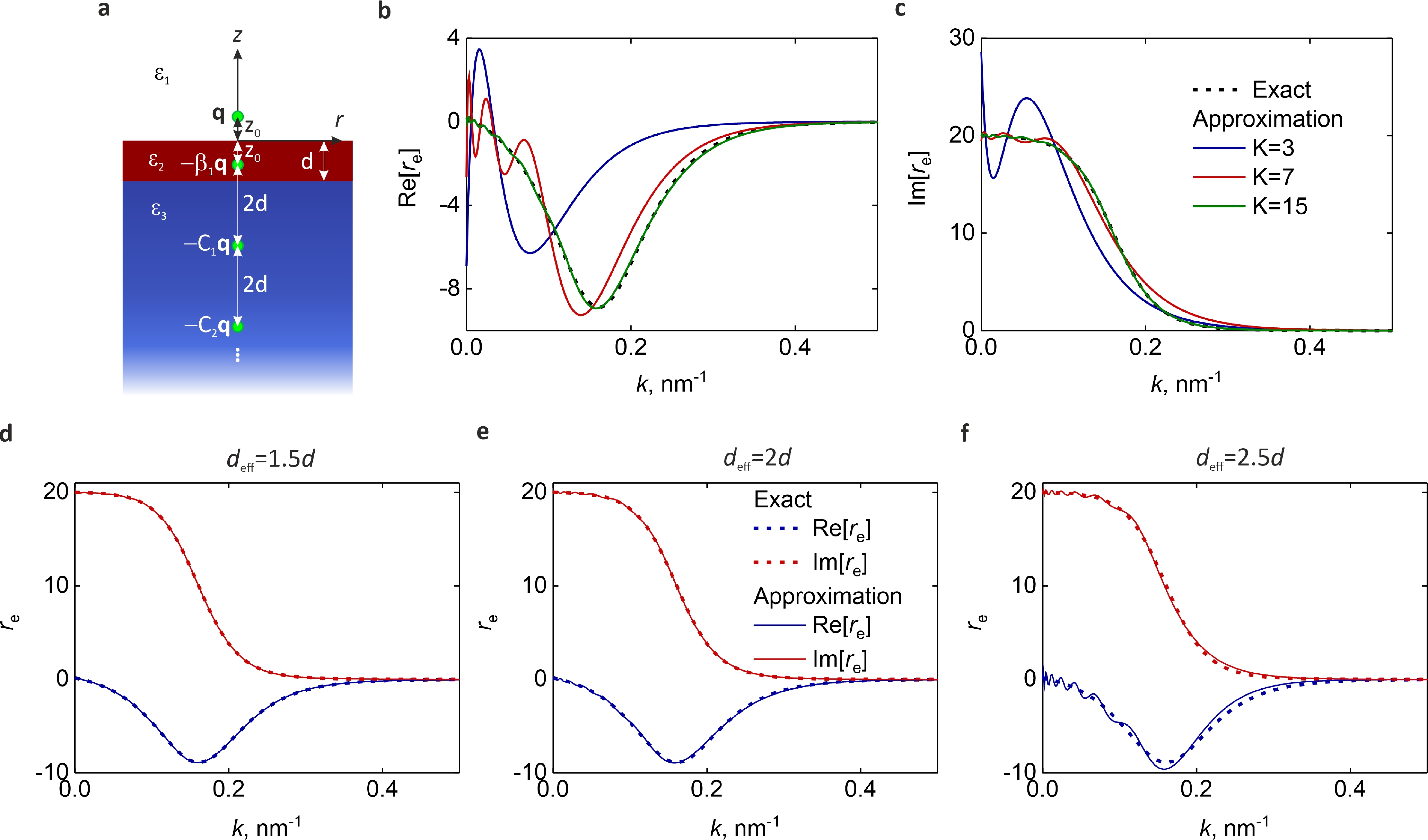}
\caption{
a) Schematic of a point charge $q$ above a layer of thickness $d$ and the corresponding image charges used to represent the electrostatic response of the layered sample.
b,c) Real (b) and imaginary (c) parts of the exact electrostatic reflection coefficient calculated using Equation \eqref{Re_layer} (dashed curve), together with polynomial approximations in $x=e^{-2k_\|d}$, Equations \eqref{ReReW} and \eqref{ImReW}, for different polynomial orders $K$.
d--f) Approximation of the electrostatic reflection coefficient using polynomials of $x=e^{-2k_\|d_{\mathrm{eff}}}$ of fixed order $K=15$ for $d_{\mathrm{eff}}=1.5d$ (d), $d_{\mathrm{eff}}=2d$ (e), and $d_{\mathrm{eff}}=2.5d$ (f).
}
\label{Reflection_fitting}
\end{figure}

To introduce the electrostatic image method, we consider a point charge $q$ located at $(\mathbf{r}_0,z_0)$ above a planar structure whose upper interface is defined by $z=0$, as illustrated in Figure \ref{Reflection_fitting}a. The electrostatic potential in the upper half-space with dielectric permittivity $\eps_1$ can be written as
\begin{equation}
\varphi(\mathbf{r},z)
=
\frac{q}{\eps_1}
\left[
\iint\limits_{\mathbb{R}^2}
\frac{d^2\mathbf{k}_\|}{2\pi k_\|}
e^{i\mathbf{k}_\|\cdot(\mathbf{r}-\mathbf{r}_0)-k_\||z-z_0|}
+
\iint\limits_{\mathbb{R}^2}
\frac{d^2\mathbf{k}_\|}{2\pi k_\|}
r_e(\mathbf{k}_\|)
e^{i\mathbf{k}_\|\cdot(\mathbf{r}-\mathbf{r}_0)-k_\|(z+z_0)}
\right],
\label{GF_r}
\end{equation}
where $\mathbf{r}$ and $z$ denote the in-plane and out-of-plane coordinates, respectively, $\mathbf{k}_\|$ is the in-plane wavevector of the corresponding Fourier harmonic of the potential, and $k_\|=|\mathbf{k}_\||$. The first term in Equation \eqref{GF_r} is the potential of the source charge, whereas the second term describes the electrostatic response of the sample. The quantity $r_e(\mathbf{k}_\|)$ is the electrostatic reflection coefficient, i.e., the reflection coefficient for spatial Fourier harmonics of the electrostatic potential. For the in-plane isotropic structures considered below, it depends only on the wavevector magnitude, $r_e(\mathbf{k}_\|)=r_e(k_\|)$.

For a homogeneous bulk material with dielectric permittivity $\eps_2$, the electrostatic reflection coefficient is independent of the in-plane wavevector and is given by
\begin{equation}
r_e(k_\|)
=
\frac{\eps_1-\eps_2}{\eps_1+\eps_2}
=
-\beta_0.
\label{Re_isotr}
\end{equation}
The Fourier integrals in Equation \eqref{GF_r} can then be evaluated analytically, yielding
\begin{equation}
\varphi(\mathbf{r},z)
=
\frac{q}{\eps_1}
\left[
\frac{1}{
\sqrt{
|\mathbf{r}-\mathbf{r}_0|^2+(z-z_0)^2
}
}
-
\frac{\beta_0}{
\sqrt{
|\mathbf{r}-\mathbf{r}_0|^2+(z+z_0)^2
}
}
\right].
\label{GF_r_i}
\end{equation}
The first term in Equation \eqref{GF_r_i} is the potential produced by the original charge $q$ located at $(\mathbf{r}_0,z_0)$, while the second term is identical to the potential produced by an image charge $-\beta_0q$ located at $(\mathbf{r}_0,-z_0)$. Thus, the electrostatic response of a homogeneous dielectric half-space can be represented exactly by a single image charge.

A similar result holds for a uniaxial material whose optical axis is perpendicular to the surface. Denoting the in-plane and out-of-plane dielectric permittivities by $\eps_{\perp}$ and $\eps_{\|}$, respectively, the electrostatic reflection coefficient is
\begin{equation}
r_e(k_\|)
=
\frac{
\eps_1-\sqrt{\eps_{\perp}\eps_{\|}}
}{
\eps_1+\sqrt{\eps_{\perp}\eps_{\|}}
}.
\label{Re_uniax}
\end{equation}
It is again independent of $k_\|$. Therefore, for a point charge $q$ located at $(0,0,z_0)$, the electrostatic response of the uniaxial half-space is represented by an image charge $-\beta_u q$ located at $(0,0,-z_0)$, where
\begin{equation}
\beta_u
=
\frac{
\sqrt{\eps_{\perp}\eps_{\|}}-\eps_1
}{
\sqrt{\eps_{\perp}\eps_{\|}}+\eps_1
}.
\label{Beta_uniax}
\end{equation}

The situation becomes more complex for a layered sample. We first consider a single isotropic layer of thickness $d$ and dielectric permittivity $\eps_2$ occupying $-d<z<0$, placed on a semi-infinite substrate with dielectric permittivity $\eps_3$. The upper half-space has dielectric permittivity $\eps_1$. The electrostatic reflection coefficient of this structure depends on the in-plane wavevector and reads
\begin{equation}
r_e(k_\|)
=
-
\frac{
\beta_1+\beta_2e^{-2k_\|d}
}{
1+\beta_1\beta_2e^{-2k_\|d}
},
\label{Re_layer}
\end{equation}
where
\begin{equation}
\beta_1
=
\frac{\eps_2-\eps_1}{\eps_2+\eps_1},
\qquad
\beta_2
=
\frac{\eps_3-\eps_2}{\eps_3+\eps_2}.
\label{Beta_layer}
\end{equation}

Because of this wavevector dependence, the reflected contribution in Equation \eqref{GF_r} can no longer be reduced directly to a single image charge. However, provided that $|\beta_1\beta_2|<1$, the denominator in Equation \eqref{Re_layer} can be expanded as a geometric series, giving
\begin{equation}
r_e(k_\|)
=
-\beta_1
-
\beta_2(1-\beta_1^2)
\sum\limits_{n=0}^{\infty}
(-\beta_1\beta_2)^n
e^{-2(n+1)k_\|d}.
\label{Re_layer_sum1}
\end{equation}
Substituting Equation \eqref{Re_layer_sum1} into Equation \eqref{GF_r}, interchanging the order of summation and integration, and evaluating the Fourier integral term by term yields
\begin{multline}
\varphi(\mathbf{r},z)
=
\frac{q}{\eps_1}
\left[
\frac{1}{
\sqrt{
|\mathbf{r}-\mathbf{r}_0|^2+(z-z_0)^2
}
}
-
\frac{\beta_1}{
\sqrt{
|\mathbf{r}-\mathbf{r}_0|^2+(z+z_0)^2
}
}
\right.
\\
\left.
-
\sum_{n=1}^{\infty}
\frac{
\beta_2(1-\beta_1^2)
(-\beta_1\beta_2)^{n-1}
}{
\sqrt{
|\mathbf{r}-\mathbf{r}_0|^2
+
(z+z_0+2nd)^2
}
}
\right].
\label{GF_r_weak_layer}
\end{multline}
Equation \eqref{GF_r_weak_layer} provides a direct image-charge representation of the electrostatic response of the layered sample. In addition to the first image charge, $q_0=-\beta_1q$, located at a depth $z_0$ below the upper interface, the response is represented by an infinite series of image charges
\begin{equation}
q_n
=
-\beta_2(1-\beta_1^2)
(-\beta_1\beta_2)^{n-1}q,
\qquad n=1,2,\ldots,
\end{equation}
located at depths $z_n=z_0+2nd$.
Thus, the wavevector-dependent response of the layer is transformed into the electrostatic field of a discrete set of image charges. This construction is the conventional generalization of the electrostatic image method to a dielectric layer \cite{Sometani_2000}.

The crucial limitation of this conventional expansion is the convergence condition $|\beta_1\beta_2|<1$. This condition can readily be violated for strongly resonant layers, particularly in spectral regions where the real part of $\eps_2$ becomes negative. To remove this restriction, we introduce a generalized image-charge method (GICM). We first introduce the variable $x=e^{-2k_\|d}$.
For a material with finite optical losses, the pole associated with the denominator of Equation \eqref{Re_layer} is shifted away from the real $k_\|$ axis, and $r_e(x)$ is finite and smooth over the interval $x\in[0,1]$. Its real and imaginary parts are therefore continuous functions on this interval and, according to the Weierstrass approximation theorem, can be approximated to arbitrary accuracy by polynomials in $x$:
\begin{equation}
\operatorname{Re}[r_e(x)]
\approx
-\operatorname{Re}[\beta_1]
-
\sum_{n=1}^{K}A_nx^n,
\label{ReReW}
\end{equation}
and
\begin{equation}
\operatorname{Im}[r_e(x)]
\approx
-\operatorname{Im}[\beta_1]
-
\sum_{n=1}^{K}B_nx^n.
\label{ImReW}
\end{equation}
Combining these two expressions gives
\begin{equation}
r_e(k_\|)
\approx
-\beta_1
-
\sum_{n=1}^{K}
C_ne^{-2nk_\|d},
\label{ReW}
\end{equation}
where
\begin{equation}
C_n=A_n+iB_n.
\end{equation}
The coefficients $C_n$ can be obtained numerically, for example by least-squares fitting of the exact reflection coefficient in Equation \eqref{Re_layer}.

The representation in Equation \eqref{ReW} again allows the Fourier integral in Equation \eqref{GF_r} to be evaluated analytically. The resulting electrostatic potential is
\begin{multline}
\varphi(\mathbf{r},z)
=
\frac{q}{\eps_1}
\left[
\frac{1}{
\sqrt{
|\mathbf{r}-\mathbf{r}_0|^2+(z-z_0)^2
}
}
-
\frac{\beta_1}{
\sqrt{
|\mathbf{r}-\mathbf{r}_0|^2+(z+z_0)^2
}
}
\right.
\\
\left.
-
\sum_{n=1}^{K}
\frac{
C_n
}{
\sqrt{
|\mathbf{r}-\mathbf{r}_0|^2
+
(z+z_0+2nd)^2
}
}
\right].
\label{GF_r_GICM}
\end{multline}
Thus, within the GICM, the electrostatic response of a single layer on a semi-infinite substrate is represented by the image charge $q_0=-\beta_1q$ at a depth $z_0$, together with $K$ additional image charges
\begin{equation}
q_n=-C_nq,
\qquad n=1,\ldots,K,
\end{equation}
located at depths $z_n=z_0+2nd$.
Unlike the conventional image-charge series in Equation \eqref{GF_r_weak_layer}, this representation is not restricted by the condition $|\beta_1\beta_2|<1$.

The approximation error of GICM can be systematically reduced by increasing the polynomial order $K$. To illustrate the convergence in a particularly demanding case, we consider a $10$-nm-thick layer supporting highly confined, low-loss modes, as shown in Figure \ref{Reflection_fitting}a. We use $\eps_2=-0.1+0.1i$ for the layer, $\eps_3=12$ for the substrate, and $\eps_1=1$ for the superstrate. The small imaginary part of $\eps_2$, together with its negative real part, produces a sharp wavevector-dependent feature in the electrostatic reflection coefficient and therefore provides a stringent test of the polynomial approximation. Figures \ref{Reflection_fitting}b,c compare the exact reflection coefficient with approximations obtained for different values of $K$. For the parameters considered here, good convergence is already achieved at $K=15$.

More generally, the expansion is not restricted to the variable $e^{-2k_\|d}$. The reflection coefficient can instead be represented as
\begin{equation}
r_e(k_\|)
\approx
-\beta_1
-
\sum_{n=1}^{K}
C_n
e^{-2nk_\|d_{\mathrm{eff}}},
\label{ReW_deff}
\end{equation}
where $d_{\mathrm{eff}}$ is an arbitrary positive length scale. In this representation, the effective image charges are separated by a distance $2d_{\mathrm{eff}}$. Although the approximation can in principle be made arbitrarily accurate for different choices of $d_{\mathrm{eff}}$, this choice strongly affects the convergence rate and therefore the number of terms $K$ required to reach a specified accuracy.

Figures \ref{Reflection_fitting}d--f illustrate this dependence for a fixed polynomial order $K=15$ and $d_{\mathrm{eff}}=1.5d$, $2d$, and $2.5d$, respectively. For the example considered here, the approximation remains accurate for $d_{\mathrm{eff}}\lesssim2d$, whereas its convergence deteriorates substantially when $d_{\mathrm{eff}}$ is increased further. This behavior is important for extending the method to general multilayer structures, where several geometrical length scales may be present. In practical calculations, $d_{\mathrm{eff}}$ should therefore be chosen sufficiently small compared with the smallest layer thickness in the structure. In the multilayer calculations presented in this work, we choose the spacing between successive image charges, $2d_{\mathrm{eff}}$, equal to the thickness of the thinnest layer in the stack.

\section{Implementation of GICM into ESM}

We now incorporate GICM into the exact spheroid model (ESM). As shown in Section~S1, the polarization charge density on the spheroid can be expanded as
\begin{equation}
\rho(\eta)
=
\frac{
\sum\limits_{n=0}^{\infty}(2n+1)\upsilon_n P_n(\eta)
}{
2\pi c\sqrt{(\xi_0^2-1)(\xi_0^2-\eta^2)}
}.
\label{Sph_sig_eps}
\end{equation}
For a homogeneous bulk sample, the electrostatic response of the sample is represented by a single image spheroid, and the corresponding sample-induced potential is given by Equations~\eqref{Sph_phi_i_sigma} and \eqref{Sph_phi_i_sigma_2_eps}. For a layered sample, GICM replaces this single image by the set of effective image spheroids shown schematically in Figure~\ref{Ellipsoid_scheme}b.

For a single layer on a semi-infinite substrate, according to GICM derived in Section~S2, the first image spheroid has the same mirror position as in the bulk problem and carries the charge density $-\beta_1\rho(\eta)$, whereas the $k^\texttt{th}$ additional image carries the charge density $-C_k\rho(\eta)$ and is displaced by an additional distance $2kd$ from the first image.

The resulting sample-induced potential at the surface of the real spheroid can therefore be written as
\begin{multline}
\varphi_s(\xi_0,\eta)
=
-\beta_1
\sum\limits_{n=0}^{\infty}
\sum\limits_{m=0}^{\infty}
(2n+1)(2m+1)
\upsilon_n
J_{nm}
P_m(\eta)P_m(\xi_0)
\\
+
\beta_1
\frac{
c(\eps-1)E_0
}{
2
\left(
\eps Q_1(\xi_0)
-\xi_0Q_0(\xi_0)
+\frac{\xi_0^2}{\xi_0^2-1}
\right)
}
\sum\limits_{m=0}^{\infty}
(2m+1)
J_{1m}
P_m(\eta)P_m(\xi_0)
\\
-
\sum\limits_{n=0}^{\infty}
\sum\limits_{m=0}^{\infty}
(2n+1)(2m+1)
\upsilon_n
\sum\limits_{k=1}^{K}
C_k J_{nm}^{k}
P_m(\eta)P_m(\xi_0)
\\
+
\frac{
c(\eps-1)E_0
}{
2
\left(
\eps Q_1(\xi_0)
-\xi_0Q_0(\xi_0)
+\frac{\xi_0^2}{\xi_0^2-1}
\right)
}
\sum\limits_{m=0}^{\infty}
(2m+1)
\sum\limits_{k=1}^{K}
C_k J_{1m}^{k}
P_m(\eta)P_m(\xi_0).
\label{Sph_phi_i_film}
\end{multline}

Here, $J_{nm}^{k}$ has the same form as the geometrical coupling integral $J_{nm}$ introduced in Equation~\eqref{Sph_int_J_sig}, but is evaluated for the $k^\texttt{th}$ effective image spheroid:
\begin{equation}
J_{nm}^{k}
=
\int\limits_{-1}^{1}
d\eta^{\prime}\,
P_n(\eta^{\prime})
P_m(\eta_q^{k})
Q_m(\xi_q^{k}).
\label{Sph_int_J_film}
\end{equation}
The spheroidal coordinates $(\xi_q^{k},\eta_q^{k})$ of the corresponding image point, expressed in the coordinate system associated with the real spheroid, are
\begin{equation}
\eta_q^{k}
=
\frac{\psi^{k}}{\xi_q^{k}},
\qquad
\xi_q^{k}
=
\sqrt{
\frac{
1+(\psi^{k})^2+\chi^2
+
\sqrt{
\left[1+(\psi^{k})^2+\chi^2\right]^2
-4(\psi^{k})^2
}
}{2}
},
\label{Sph_eta_xi_q_film}
\end{equation}
where
\begin{equation}
\psi^{k}
=
2\frac{a+H+kd}{c}
-\xi_0\eta^{\prime},
\qquad
\chi
=
\sqrt{
(\xi_0^2-1)(1-\eta^{\prime 2})
}.
\label{Sph_psi_chi_q_film}
\end{equation}

The remaining derivation is identical to that for a homogeneous bulk sample. Indeed, comparison of Equation~\eqref{Sph_phi_i_film} with Equation~\eqref{Sph_phi_i_sigma_2_eps} shows that the effect of the layered sample can be incorporated by replacing every occurrence of the bulk coupling term $\beta J_{nm}$ according to
\begin{equation}
\beta J_{nm}
\;\longrightarrow\;
\beta_1 J_{nm}
+
\sum_{k=1}^{K}
C_k J_{nm}^{k}.
\label{Sph_int_J_gen_film}
\end{equation}
Thus, GICM modifies only the electrostatic coupling between the spheroid and the sample, while the structure of the boundary-value problem derived in Section~S1 remains unchanged.

The vertical polarizability of the spheroid above the layered sample therefore retains the same form as in the bulk case:
\begin{equation}
\alpha_z
=
2ac\vartheta_1^{l}
-
\frac{
(\eps-1)c^3P_1\left(\frac{a}{c}\right)
}{
3
\left[
\eps Q_1\left(\frac{a}{c}\right)
-
P_1\left(\frac{a}{c}\right)
Q_0\left(\frac{a}{c}\right)
+
\frac{a^2}{a^2-c^2}
\right]
},
\label{Sph_alphaZ_eps_sig_l}
\end{equation}
where the superscript $l$ denotes the layered-sample solution. The coefficients $\vartheta_n^l=\upsilon_n^l/E_0$ are obtained from the linear system
\begin{equation}
\sum\limits_{n=1}^{\infty}
\mathcal{M}_{mn}^l
\vartheta_n^l
=
\mathcal{C}_m^l,
\qquad m\geq1,
\label{Sph_syst_eps_sig_l}
\end{equation}
with
\begin{equation} \mathcal{M}_{m n}^l = \delta_{nm} - \frac{(\eps-1) (2 n + 1) \left[ \beta_1 J_{n m} + \sum_{k=1}^K C_k J_{n m}^k \right]}{2 P_m(\xi_0) \left(\eps \frac{Q_m(\xi_0)}{P_m(\xi_0)} - \frac{\xi_0 Q_m(\xi_0) - Q_{m-1}(\xi_0)}{\xi_0 P_m(\xi_0) - P_{m-1}(\xi_0)}\right)}, \label{Sph_Mat_eps_sig1} \end{equation} \begin{equation} \mathcal{C}_m^l = - \frac{c (\eps-1)^2 \left[ \beta_1 J_{1 m} + \sum_{k=1}^K C_k J_{1 m}^k \right]}{4 P_m(\xi_0) \left( \eps Q_1(\xi_0) - \xi_0 Q_0(\xi_0) + \frac{\xi_0^2}{\xi_0^2-1} \right) \left(\eps \frac{Q_m(\xi_0)}{P_m(\xi_0)} - \frac{\xi_0 Q_m(\xi_0) - Q_{m-1}(\xi_0)}{\xi_0 P_m(\xi_0) - P_{m-1}(\xi_0)}\right)}. \label{Sph_C_eps_sig1} \end{equation}
As in the bulk case, the infinite system is truncated at a sufficiently large multipole order, determined by convergence of the calculated polarizability. The only additional numerical step in the multilayer formulation is therefore the evaluation of the coupling integrals $J_{nm}^{k}$ for the effective image spheroids introduced by GICM.

\section{Implementation of FIM in ESM}

As an alternative to GICM, we incorporate the response of a layered sample into ESM directly through the Fourier representation of its electrostatic reflection coefficient. We call this approach the Fourier integral method (FIM). In contrast to GICM, where $r_e(k_\|)$ is represented by a discrete set of effective image charges, FIM retains the Fourier integral in Equation~\eqref{GF_r} and evaluates the sample-induced potential by integrating the reflected contribution from every surface charge element of the spheroid.

We assume that the sample is isotropic in the surface plane, such that $r_e(\mathbf{k}_\|)=r_e(k_\|)$, and, as in the preceding derivation of ESM, take the upper half-space to be vacuum. The potential induced by the sample due to the complete surface charge distribution $\rho(\eta)$ on the spheroid can then be written as
\begin{multline}
\varphi_s(\xi,\eta)
=
\int\limits_{-1}^{1}
\int\limits_{0}^{2\pi}
\rho(\eta^\prime)c^2
\sqrt{(\xi_0^2-1)(\xi_0^2-\eta^{\prime 2})}
\,d\eta^\prime d\phi^\prime
\\
\times
\int\limits_{0}^{\infty}
\int\limits_{0}^{2\pi}
\frac{dk_\|\,d\theta}{2\pi}
r_e(k_\|)
e^{ik_\|r\cos(\theta-\phi)}
e^{-ik_\|r^\prime\cos(\theta-\phi^\prime)}
e^{-k_\|(z+z^\prime)},
\label{Sample_pot}
\end{multline}
where $\theta$ is the polar angle of the in-plane wavevector $\mathbf{k}_\|$. Here, $(r,\phi,z)$ are cylindrical coordinates referred to the sample surface as shown in Figure \ref{Reflection_fitting}a, while $(r^\prime,\phi^\prime,z^\prime)$ denote the coordinates of a surface element of the spheroid with spheroidal coordinates $(\xi_0,\eta^\prime,\phi^\prime)$, which are introduced in Figure \ref{Ellipsoid_scheme}. The cylindrical and spheroidal coordinate systems share the same azimuthal angle, and their radial and vertical coordinates are related by
\begin{equation}
\begin{array}{l}
r=c\sqrt{(\xi^2-1)(1-\eta^2)},\\[3pt]
z=a+H-c\xi\eta.
\end{array}
\label{Sph_cyl_coord}
\end{equation}
For a source point on the spheroid surface, the corresponding relations are obtained by setting $\xi=\xi_0$ and $\eta=\eta^\prime$.

We now decompose the spheroid charge density as
$\rho(\eta)=\rho_i(\eta)+\rho_0(\eta)$, where the interaction-induced contribution $\rho_i$ is expanded according to Equation~\eqref{Sph_sig}, while $\rho_0$ is given by Equation~\eqref{Sph_sig_0_eps}. Substituting these expressions into Equation~\eqref{Sample_pot} and carrying out the integrations over $\phi^\prime$ and $\theta$ gives
\begin{multline}
\varphi_s(\xi,\eta)
=
c\sum\limits_{n=0}^{\infty}(2n+1)\upsilon_n
\int\limits_{0}^{\infty}
dk_\|\,r_e(k_\|)
\int\limits_{-1}^{1}
d\eta^\prime\,
P_n(\eta^\prime)
J_0(k_\|r)
J_0(k_\|r^\prime)
e^{-k_\|(z+z^\prime)}
\\
-
\frac{(\eps-1)E_0c^2}
{2\left(
\eps Q_1(\xi_0)
-\xi_0Q_0(\xi_0)
+\frac{\xi_0^2}{\xi_0^2-1}
\right)}
\int\limits_{0}^{\infty}
dk_\|\,r_e(k_\|)
\int\limits_{-1}^{1}
d\eta^\prime\,
\eta^\prime
J_0(k_\|r)
J_0(k_\|r^\prime)
e^{-k_\|(z+z^\prime)},
\label{Sample_pot_cylindr}
\end{multline}
where $J_0$ is the Bessel function of the first kind.

The integrals over $\eta^\prime$ can be evaluated using the identity \cite{Prudnikov}
\begin{equation}
\int\limits_{-1}^{1}
d\eta^\prime\,
P_n(\eta^\prime)
J_0\left(
k_\|c\sqrt{(\xi_0^2-1)(1-\eta^{\prime 2})}
\right)
e^{k_\|c\xi_0\eta^\prime}
=
\sqrt{\frac{2\pi}{k_\|c}}
P_n(\xi_0)
I_{n+\frac{1}{2}}(k_\|c),
\label{Integral_calc}
\end{equation}
where $I_\nu$ is the modified Bessel function of the first kind. Applying Equation~\eqref{Integral_calc} to Equation~\eqref{Sample_pot_cylindr}, and using $P_1(\xi_0)=\xi_0=a/c$ for the contribution proportional to $\rho_0$, yields
\begin{multline}
\varphi_s(\xi,\eta)
=
c\sum\limits_{n=0}^{\infty}
(2n+1)\upsilon_n
\int\limits_{0}^{\infty}
dk_\|\,
r_e(k_\|)
\sqrt{\frac{2\pi}{k_\|c}}
P_n(\xi_0)
I_{n+\frac{1}{2}}(k_\|c)
e^{-2k_\|(a+H)}
J_0(k_\|r)
e^{k_\|c\xi\eta}
\\
-
\frac{(\eps-1)E_0ac}
{2\left(
\eps Q_1(\xi_0)
-\xi_0Q_0(\xi_0)
+\frac{\xi_0^2}{\xi_0^2-1}
\right)}
\int\limits_{0}^{\infty}
dk_\|\,
r_e(k_\|)
\sqrt{\frac{2\pi}{k_\|c}}
I_{\frac{3}{2}}(k_\|c)
e^{-2k_\|(a+H)}
J_0(k_\|r)
e^{k_\|c\xi\eta}.
\label{Sample_pot_int}
\end{multline}

To incorporate this potential into the spheroid boundary conditions, we expand $\varphi_s(\xi_0,\eta)$ in Legendre polynomials. Using their orthogonality and applying Equation~\eqref{Integral_calc} once more gives
\begin{multline}
\varphi_s(\xi_0,\eta)
=
\pi
\sum\limits_{n=0}^{\infty}
\sum\limits_{m=0}^{\infty}
(2n+1)(2m+1)
\upsilon_n
P_m(\eta)
P_n(\xi_0)
P_m(\xi_0)
\\
\times
\int\limits_{0}^{\infty}
\frac{dk_\|}{k_\|}
r_e(k_\|)
I_{n+\frac{1}{2}}(k_\|c)
I_{m+\frac{1}{2}}(k_\|c)
e^{-2k_\|(a+H)}
\\
-
\frac{\pi(\eps-1)E_0a}
{2\left(
\eps Q_1(\xi_0)
-\xi_0Q_0(\xi_0)
+\frac{\xi_0^2}{\xi_0^2-1}
\right)}
\sum\limits_{m=0}^{\infty}
(2m+1)
P_m(\eta)
P_m(\xi_0)
\\
\times
\int\limits_{0}^{\infty}
\frac{dk_\|}{k_\|}
r_e(k_\|)
I_{\frac{3}{2}}(k_\|c)
I_{m+\frac{1}{2}}(k_\|c)
e^{-2k_\|(a+H)}.
\label{Sample_pot_series}
\end{multline}

The remaining derivation is identical to that for a homogeneous bulk sample. Comparing Equation~\eqref{Sample_pot_series} with the bulk expression in Equation~\eqref{Sph_phi_i_sigma_2_eps}, the complete effect of the sample can be incorporated by replacing the bulk coupling term $\beta J_{nm}$ according to
\begin{equation}
\beta J_{nm}
\;\longrightarrow\;
\mathcal{J}_{nm}
=
-\pi P_n(\xi_0)
\int\limits_{0}^{\infty}
\frac{dk_\|}{k_\|}
r_e(k_\|)
I_{n+\frac{1}{2}}(k_\|c)
I_{m+\frac{1}{2}}(k_\|c)
e^{-2k_\|(a+H)}.
\label{Sph_int_J_FIM}
\end{equation}
Thus, FIM preserves the structure of the ESM linear system and modifies only the sample--spheroid coupling term. Unlike GICM, it uses the electrostatic reflection coefficient $r_e(k_\|)$ directly and therefore does not require its intermediate representation as a finite set of image charges. However, this approach requires numerically integrating Equation \eqref{Sph_int_J_FIM}.

We next describe the numerical evaluation of Equation~\eqref{Sph_int_J_FIM}. In principle, the integral can be evaluated by direct quadrature along the real $k_\parallel$ axis. This approach, however, becomes inefficient for layers supporting weakly damped guided modes. The electrostatic reflection coefficient $r_e(k_\parallel)$ is meromorphic in $k_\parallel$ in the whole complex plane, with poles corresponding to the electrostatic guided modes of the layered sample. For low optical losses, some of these poles approach the real axis and produce narrow resonant features in the integrand. Their positions vary with frequency, layer thickness, and the sample's optical properties. Consequently, a quadrature grid capable of resolving these features must generally be adapted for each frequency, and the required number of integration points can increase substantially for thicker low-loss layers.

This difficulty can be avoided by deforming the integration contour into the complex $k_\parallel$ plane. We introduce the kernel
\begin{equation}
\mathcal{K}_{nm}(k_\parallel,H)
=
-\frac{\pi}{k_\parallel}
P_n(\xi_0)
I_{n+\frac{1}{2}}(k_\parallel c)
I_{m+\frac{1}{2}}(k_\parallel c)
e^{-2k_\parallel(a+H)}.
\label{eq:FIM_kernel}
\end{equation}
Although Equation~\eqref{eq:FIM_kernel} contains an explicit factor $1/k_\parallel$, the apparent singularity at $k_\parallel=0$ is removable because the product of the modified Bessel functions scales as $k_\parallel^{n+m+1}$. The kernel is therefore an analytic function of $k_\parallel$. We deform the positive real axis to a ray
\begin{equation}
k_\parallel=\kappa e^{i\sigma\theta},
\qquad
\kappa\geq0,
\qquad
\sigma=\pm1,
\qquad
0<\theta<\frac{\pi}{2},
\end{equation}
where $\sigma$ selects the direction of the deformation into the upper or lower half-plane. Provided that no poles of $r_e(k_\parallel)$ lie in the sector between the original and deformed contours, Cauchy's theorem gives
\begin{equation}
\mathcal{J}_{nm}
=
\int_0^\infty
r_e(k_\parallel)
\mathcal{K}_{nm}(k_\parallel,H)
\,dk_\parallel
=
e^{i\sigma\theta}
\int_0^\infty
r_e\!\left(\kappa e^{i\sigma\theta}\right)
\mathcal{K}_{nm}\!\left(\kappa e^{i\sigma\theta},H\right)
\,d\kappa.
\label{eq:S67}
\end{equation}
The contribution from the arc closing the contour at large $|k_\parallel|$ vanishes because the kernel decays exponentially, as $\left|\mathcal{K}_{nm}\right| \lesssim e^{-2(a+H-c)\operatorname{Re}k_\parallel}$, with $a>c$. Equation~\eqref{eq:S67} is therefore an exact contour deformation: only the integration path changes, while the integral's value remains unchanged.

The practical advantage of this deformation is that the ray can be chosen to remain well separated from the guided-mode poles closest to the positive real axis. The narrow resonant features encountered on the real axis are then absent, and the integrand becomes a smooth function of $\kappa$ over a broad range of frequencies and layer thicknesses. Consequently, a single fixed quadrature grid can be used for the complete spectrum, in contrast to real-axis integration, where the grid would have to resolve resonances whose positions vary with frequency.

We therefore evaluate Equation~\eqref{eq:S67} using a fixed quadrature rule on the complex ray, truncated at a maximum momentum $\kappa_{\max}$ beyond which the contribution of the kernel is negligible. Let
\begin{equation}
k_q=\kappa_q e^{i\sigma\theta},
\qquad q=1,\ldots,N_q,
\end{equation}
denote the quadrature nodes, and let $w_q$ be the corresponding weights, including the Jacobian factor $e^{i\sigma\theta}$. The coupling integral is then approximated as
\begin{equation}
\mathcal{J}_{nm}(H,\omega)
\simeq
\sum_{q=1}^{N_q}
w_q
r_e(k_q,\omega)
\mathcal{K}_{nm}(k_q,H).
\label{eq:S68}
\end{equation}
For a fixed spheroid geometry and integration contour, the nodes $k_q$ are independent of frequency and of the optical properties of the sample. All frequency and sample dependence enters Equation~\eqref{eq:S68} through the values of the reflection coefficient $r_e(k_q,\omega)$. The kernel values $\mathcal{K}_{nm}(k_q,H)$ can therefore be precomputed for all multipole indices, all tip--sample distances used to sample the tapping cycle, and all quadrature nodes.

The resulting geometry-dependent kernel matrix has a low numerical rank because $\mathcal{K}_{nm}(k_\parallel,H)$ varies smoothly with $k_\parallel$, so that the kernel values at neighboring quadrature nodes are strongly correlated. We exploit this property using an interpolative decomposition \cite{cheng2005}. The decomposition selects a subset of $N_s\ll N_q$ quadrature nodes, which we denote as the skeleton momenta $k_s$, $s=1,\ldots,N_s$, such that the kernel at every quadrature node can be approximated as
\begin{equation}
\mathcal{K}_{nm}(k_q,H)
\simeq
\sum_{s=1}^{N_s}
T_{sq}\,
\mathcal{K}_{nm}(k_s,H),
\end{equation}
where $T_{sq}$ is the interpolation matrix. Substituting this representation into Equation~\eqref{eq:S68} gives
\begin{equation}
\mathcal{J}_{nm}(H,\omega)
\simeq
\sum_{s=1}^{N_s}
\mu_s(\omega)
\mathcal{K}_{nm}(k_s,H),
\qquad
\mu_s(\omega)
=
\sum_{q=1}^{N_q}
T_{sq}\,
w_q\,
r_e(k_q,\omega).
\label{eq:S69}
\end{equation}
Thus, the expensive geometry-dependent part of the momentum integral is reduced to a sum over only $N_s$ skeleton momenta. Changing the sample therefore requires only reevaluating $r_e(k_q,\omega)$ on the fixed quadrature grid, without recalculating the spheroidal coupling kernels.

An additional advantage of this formulation is that the reflection coefficient enters only through its values at complex momenta $k_q$. Consequently, arbitrary planar multilayer structures can be treated directly through their corresponding $r_e(k_\parallel,\omega)$, including structures containing anisotropic or hyperbolic layers with complex out-of-plane propagation constants. Moreover, because the complex contour avoids the increasingly sharp real-axis resonances of thick, weakly damped layers, the number of quadrature nodes required for convergence does not increase substantially with layer thickness. Combined with the low-rank compression of the kernel matrix, this makes the FIM evaluation of Equation~\eqref{Sph_int_J_FIM} efficient even for strongly resonant multilayer structures.

\section[Relationship between FIM and GICM: coupling integrals]{Relationship between FIM and GICM and analytical evaluation of the coupling integrals}

The FIM and GICM formulations describe the same electrostatic sample response in two complementary representations. FIM retains the continuous Fourier decomposition of the reflected potential, whereas GICM approximates the same reflection coefficient by a discrete series of exponential terms corresponding to effective image charges. Consequently, substituting the GICM representation of $r_e(k_\|)$ into the FIM expression must reproduce the coupling terms obtained directly from the image-charge construction.

For a single layer, GICM represents the electrostatic reflection coefficient as
\begin{equation}
r_e(k_\|)
\approx
-\beta_1
-
\sum_{j=1}^{K}
C_j e^{-2jk_\|d}.
\end{equation}
Substituting this expression into Equation~\eqref{Sph_int_J_FIM} gives
\begin{multline}
\beta_1J_{nm}
+
\sum_{j=1}^{K}
C_jJ_{nm}^{j}
=
\pi P_n(\xi_0)
\int\limits_{0}^{\infty}
\frac{dk_\|}{k_\|}
\left(
\beta_1
+
\sum_{j=1}^{K}
C_j e^{-2jk_\|d}
\right)
\\
\times
I_{n+\frac{1}{2}}(k_\|c)
I_{m+\frac{1}{2}}(k_\|c)
e^{-2k_\|(a+H)}.
\label{FIM_GICM}
\end{multline}
Comparing the coefficients of $\beta_1$ and $C_j$ on both sides yields an alternative representation of the geometrical coupling integrals:
\begin{equation}
J_{nm}^{j}
=
\pi P_n(\xi_0)
\int\limits_{0}^{\infty}
\frac{dk_\|}{k_\|}
I_{n+\frac{1}{2}}(k_\|c)
I_{m+\frac{1}{2}}(k_\|c)
e^{-2k_\|(a+H+jd)},
\label{Jk_int}
\end{equation}
with $J_{nm}=J_{nm}^{0}$. Equation~\eqref{Jk_int} establishes the equivalence between the original geometrical integral in spheroidal coordinates, Equation~\eqref{Sph_int_J_film}, and a one-dimensional Fourier--Bessel integral. In addition to clarifying the relationship between FIM and GICM, this representation provides a convenient alternative way of evaluating the coupling coefficients.

The integral in Equation~\eqref{Jk_int} can be evaluated analytically using the corresponding integral of a product of modified Bessel functions \cite{Prudnikov}. This gives
\begin{multline}
J_{nm}^{j}
=
\pi P_n(\xi_0)
\left(
\frac{c}{4(a+H+jd)}
\right)^{n+m+1}
\frac{
\Gamma(n+m+1)
}{
\Gamma\left(n+\frac{3}{2}\right)
\Gamma\left(m+\frac{3}{2}\right)
}
\\
\times
{}_4F_3
\left(
\begin{matrix}
\frac{n+m}{2}+1,\;
\frac{n+m}{2}+\frac{3}{2},\;
\frac{n+m}{2}+\frac{1}{2},\;
\frac{n+m}{2}+1
\\[3pt]
n+m+2,\;
n+\frac{3}{2},\;
m+\frac{3}{2}
\end{matrix}
;
\frac{c^2}{(a+H+jd)^2}
\right),
\label{J_int_an}
\end{multline}
where $\Gamma$ is the Gamma function and ${}_4F_3$ is the generalized hypergeometric function. Equation~\eqref{J_int_an} therefore provides a closed-form expression for the coupling between the $n^{\text{th}}$ multipole component of the real spheroid and the $m^{\text{th}}$ multipole component associated with the $j^{\text{th}}$ image spheroid.

\section{Finite-element method simulation}

We perform finite-element method (FEM) simulations following the same procedure as in our previous work \cite{ESM}, extending the model to include a finite-thickness layer on top of the substrate. Because the system is rotationally symmetric, we solve the electrostatic problem in a two-dimensional, axially symmetric geometry. Within each homogeneous region, the potential satisfies Laplace's equation. We introduced a uniform vertical external electric field by prescribing the electric potential at the exterior boundary of the simulation domain. For the layered geometry, we modify this boundary potential relative to the bulk-sample case to reproduce the background electrostatic field of the corresponding planar multilayer structure. The potential imposed along the exterior boundary was therefore defined piecewise according to the dielectric permittivity of the superstrate, layer, and substrate, such that the resulting background solution satisfies continuity of the electrostatic potential and of the normal component of the electric displacement field at the material interfaces.

The spheroid perturbs this background field and acquires a surface charge density. The corresponding vertical dipole moment is obtained by integrating the induced surface charge density over the spheroid surface,
\begin{equation}
p_z=\int\limits_S z\,\rho(\mathbf{r})\,dS.
\end{equation}
We then calculate the spheroid polarizability as $\alpha_z=p_z/E_0$. To simulate the tapping motion of the s-SNOM tip, we solve the electrostatic problem for a series of spheroid--sample distances covering one complete tapping cycle. We then Fourier-transform the resulting distance-dependent dipole moment with respect to the tapping phase to obtain the required demodulation harmonics of the spheroid polarizability.

The computational mesh is most refined in the tip--sample gap, where the electrostatic field shows the largest spatial gradients. We set the minimum mesh size to $0.2$~nm at the point of closest approach between the spheroid apex and the sample surface and increase it away from the spheroid surface using a mesh growth rate of $1.1$. Aside from adding the layer and modifying the exterior boundary condition accordingly, all simulation procedures are identical to those used for the homogeneous bulk sample in our previous work \cite{ESM}.

\section{Additional tests of MESM for different permittivities and layer thicknesses}

To assess the accuracy and convergence of the multilayer exact spheroid model (MESM) over a broad range of optical responses, we first consider a generic layer whose dielectric permittivity is varied independently over the complex plane rather than being restricted to a particular material. The layer has a thickness of $d=20$~nm and is placed on a Si substrate with $\eps_s=12$. The spheroid parameters are $R=30$~nm, $L=600$~nm, $A=50$~nm, and $H_{\min}=3$~nm. We vary the layer permittivity over the ranges
\begin{equation}
\operatorname{Re}(\eps_l)\in[-20,20],
\qquad
\operatorname{Im}(\eps_l)\in[0,25].
\end{equation}

\begin{figure}[H]
\centering
\includegraphics[width=16cm]{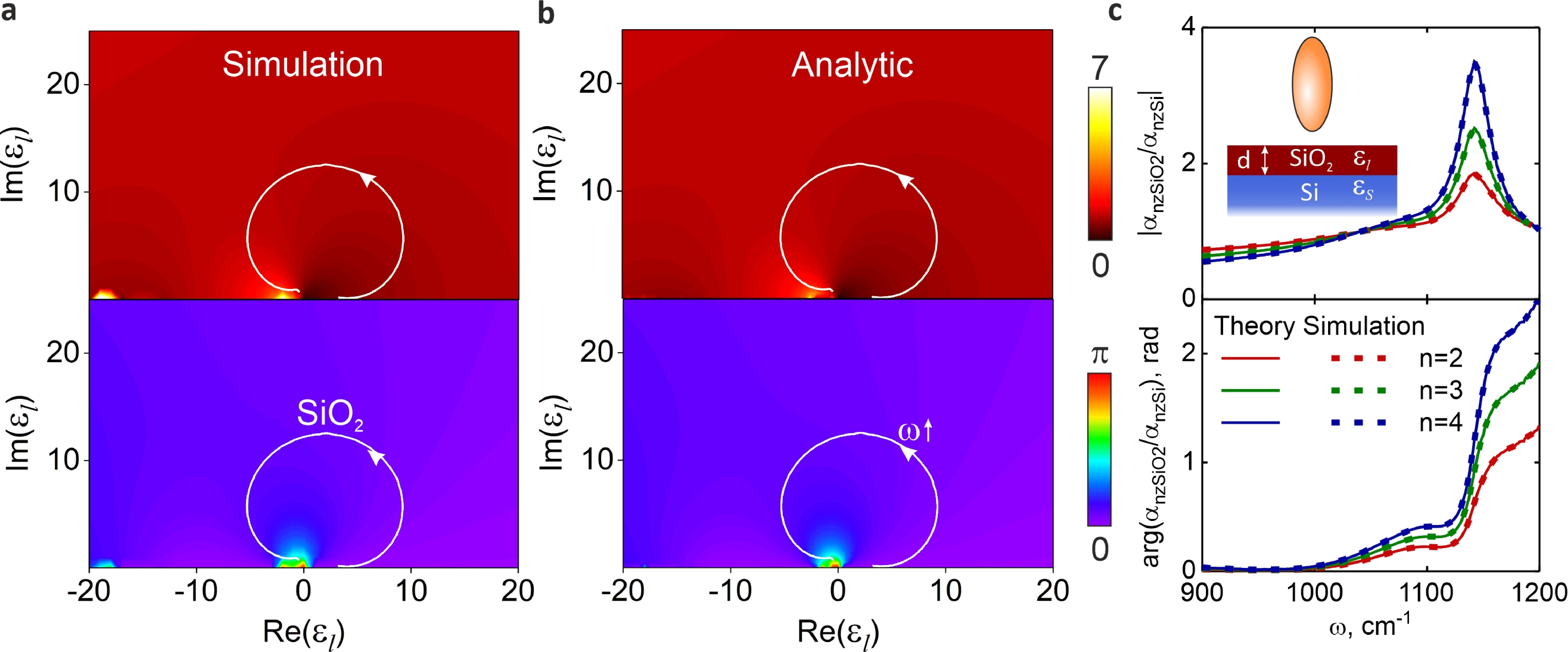}
\caption{
a) FEM-calculated amplitude (top) and phase (bottom) of the third harmonic of the spheroid polarizability above a $20$-nm-thick layer on a Si substrate with $\eps_s=12$, normalized to the corresponding polarizability above bulk Si, as a function of the real and imaginary parts of the layer dielectric permittivity. The white curve shows the dielectric permittivity of SiO$_2$ between $900$ and $1200$~cm$^{-1}$.
b) Same as in panel (a), calculated analytically using MESM with $K=10$.
c) Amplitude (top) and phase (bottom) of the second, third, and fourth harmonics of the spheroid polarizability above a $20$-nm-thick SiO$_2$ layer on Si, normalized to the corresponding response above bulk Si. Solid curves show the MESM results and dashed curves show the FEM simulations. Inset: schematic of the spheroid and layered sample. The spheroid parameters are $R=30$~nm, $L=600$~nm, $A=50$~nm, and $H_{\min}=3$~nm.
}
\label{Eps_map}
\end{figure}

Figures~\ref{Eps_map}a,b show the amplitude and phase of the third harmonic of the spheroid polarizability, normalized to the corresponding response above bulk Si, obtained using FEM simulations and MESM, respectively. For the GICM expansion order $K=10$, the analytical and numerical results agree well over nearly the entire range of dielectric permittivities considered. Noticeable deviations appear only in the region with $\operatorname{Re}(\eps)<0$ and $\operatorname{Im}(\eps)\lesssim0.1$. In this regime, the layer can support highly confined, weakly damped polaritonic modes, which produce sharp features in the wavevector-dependent electrostatic reflection coefficient and therefore require more image charges for convergence. The required value of $K$ depends on the real and imaginary parts of the dielectric permittivity.

For experimentally relevant resonant materials, the losses are generally sufficient for substantially faster convergence. To illustrate this, the white curve in Figures~\ref{Eps_map}a,b shows the dielectric permittivity of SiO$_2$ in the spectral range from $900$ to $1200$~cm$^{-1}$, encompassing its phonon resonance. Figure~\ref{Eps_map}c compares the corresponding second-, third-, and fourth-harmonic spectra calculated using MESM with the FEM results for a $20$-nm-thick SiO$_2$ layer on Si. All three harmonics agree well, showing that $K=10$ is sufficient to reproduce the strongly frequency-dependent near-field response in this case.

\begin{figure}[H]
\centering
\includegraphics[width=16cm]{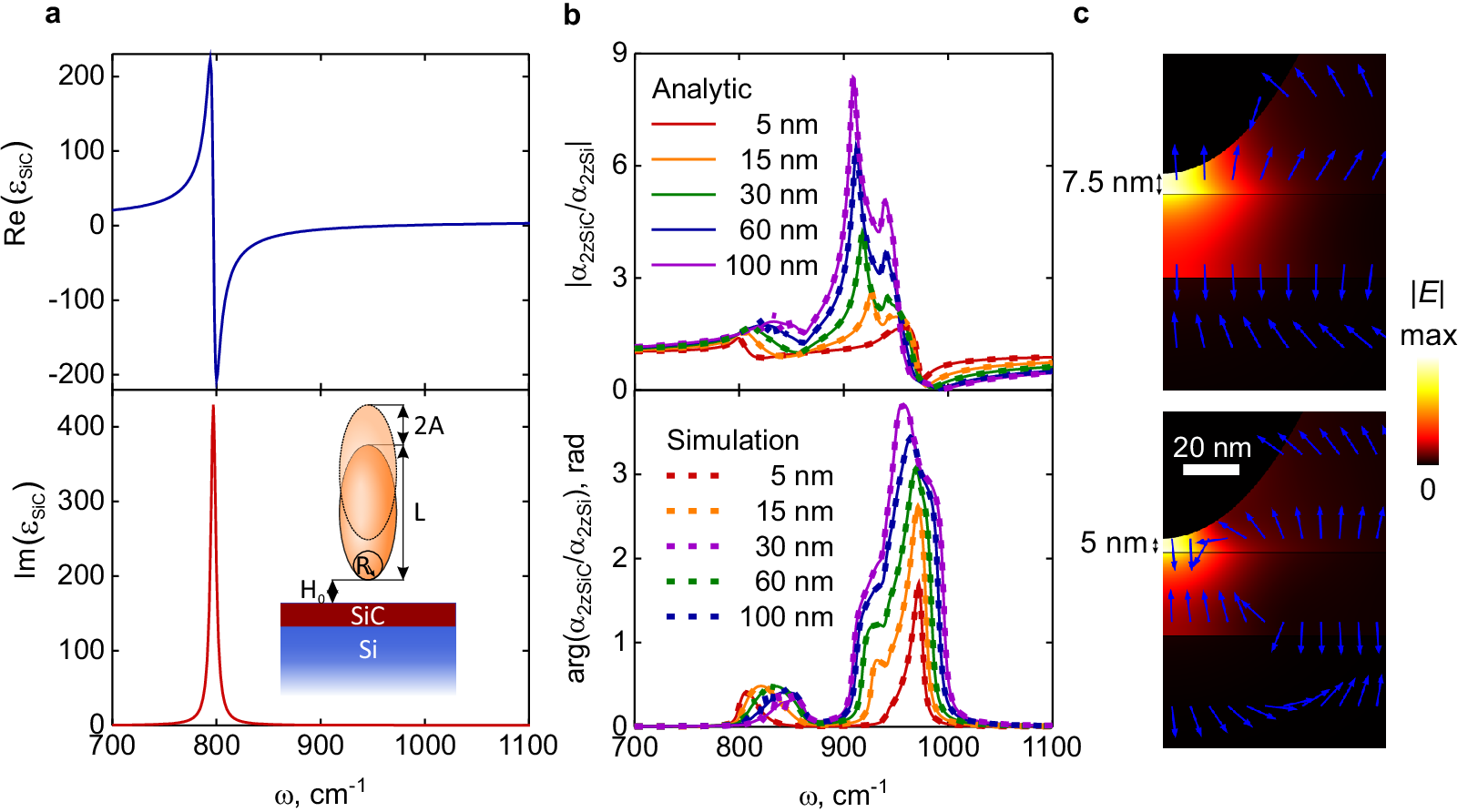}
\caption{
a) Dielectric permittivity of SiC used in the calculations. Inset: schematic of the spheroid above a SiC layer of thickness $d$ on a Si substrate, with the model's geometrical parameters.
b) Amplitude (top) and phase (bottom) of the second harmonic of the spheroid polarizability above the SiC/Si sample for layer thicknesses $d$ ranging from $5$ to $100$~nm, normalized to the corresponding response above bulk Si. Solid curves show the analytical MESM results calculated with $K=20$, and dashed curves show the FEM simulations. c) Field distribution near the apex of the tip placed above a 30-nm-thick SiC layer. Color represents the electric-field magnitude; arrows show the field direction. Top: tip-sample distance is 7.5 nm. Bottom: tip-sample distance is 5 nm.
}
\label{SiC_layers}
\end{figure}

We next examine in more detail the particularly demanding regime of low-loss materials with negative dielectric permittivity by considering 3CSiC layers of different thicknesses on a Si substrate. Figure~\ref {SiC_layers}a shows the dielectric permittivity of SiC used in the calculations. Figure~\ref{SiC_layers}b compares the second-harmonic spectra obtained using MESM and FEM for layer thicknesses ranging from $5$ to $100$~nm. The calculated spectra exhibit two main resonant features associated with the two coupled polaritonic modes supported by the SiC layer. These modes arise from coupling between the surface modes at the layer's upper and lower interfaces \cite{10.1021/acsphotonics.2c01270}. As the layer thickness increases, the higher-frequency resonance becomes more pronounced.

The main peak in the near-field spectrum clearly splits into two distinct peaks. This splitting depends on the minimum tip-sample distance and is associated with the emergence of several modes between them \cite{ESM, allen2026}. Figure \ref{SiC_layers}c shows two such modes arising between the tip and a 30 nm SiC layer at 940 cm$^{-1}$. The fundamental mode, shown at the top, appears at a distance of 7.5 nm. It is characterized by a strong field enhancement in the gap, with the field direction aligned with the external field. The first-order mode, shown at the bottom, appears at a gap of 5 nm and has a field directed opposite to the incident field. Different field distributions correspond to different tip polarizations, and the transition between them during tip oscillations also appears in the demodulated signal.

Because SiC supports highly confined and weakly damped phonon polaritons, the GICM expansion converges more slowly than in the SiO$_2$ example considered above. We therefore increased the expansion order to $K=20$. With this value, convergence is achieved for all investigated layer thicknesses, including the thinnest $5$-nm layer, and the MESM spectra reproduce the FEM calculations with excellent accuracy over the entire spectral range. Moreover, for thinner layers, convergence was achieved at $K=7-10$. These results confirm that the GICM expansion remains applicable to thin layers of strongly resonant, low-loss polaritonic materials, provided that a sufficiently large number of image charges is retained.

\section{Application of MESM to multilayer structures}

To demonstrate that the MESM is not restricted to a single layer on a semi-infinite substrate, we consider a three-component SiO$_2$/SiC/Si structure. The sample consists of a $10$-nm-thick SiO$_2$ layer deposited on a $20$-nm-thick SiC layer supported by a semi-infinite Si substrate. The dielectric permittivity of SiO$_2$ matches that used in Figure~\ref{Eps_map}, while the dielectric permittivity of SiC is taken from Figure~\ref{SiC_layers}. The spheroid parameters are $R=30$~nm, $L=600$~nm, $H_{\min}=2$~nm, and $A=50$~nm. The calculated spectra are normalized to the corresponding response above bulk Si.

\begin{figure}[H]
\centering
\includegraphics[width=16cm]{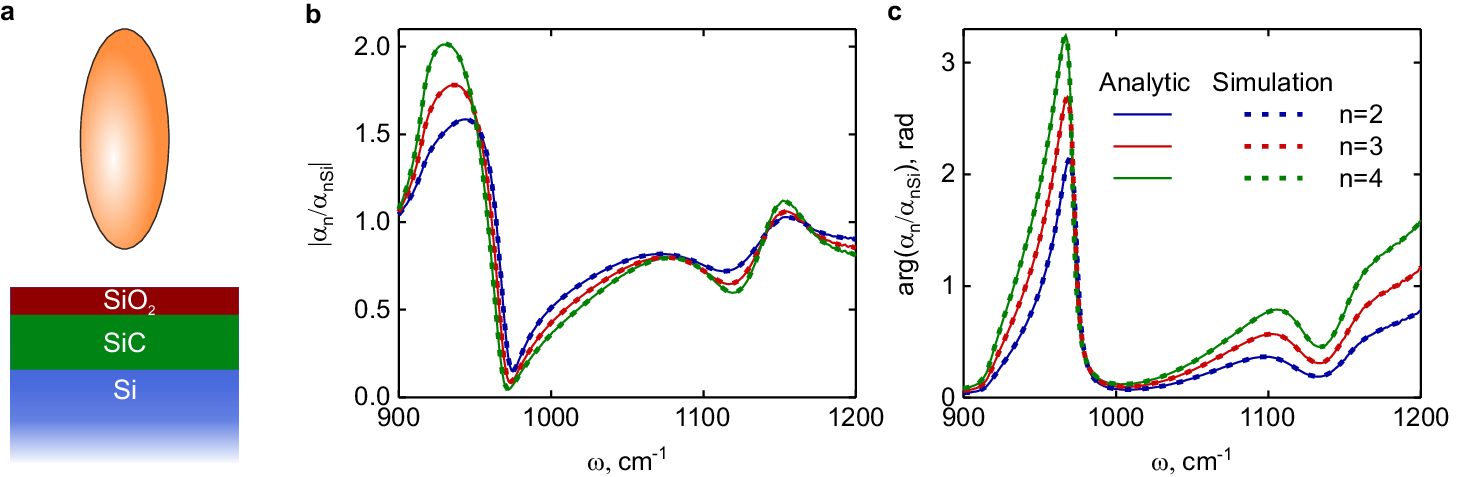}
\caption{
a) Schematic of the spheroid above a SiO$_2$/SiC/Si multilayer structure consisting of a $10$-nm-thick SiO$_2$ layer and a $20$-nm-thick SiC layer on a semi-infinite Si substrate. The spheroid parameters are $R=30$~nm, $L=600$~nm, $H_{\min}=2$~nm, and $A=50$~nm.
b,c) Amplitude (b) and phase (c) of the second, third, and fourth harmonics of the polarizability of the spheroid placed above the SiO$_2$/SiC/Si structure. The spectra are normalized to the corresponding response above bulk Si. Solid curves show the analytical MESM results and dashed curves show the FEM simulations.}
\label{SiO2_SiC_Si}
\end{figure}

Figure~\ref{SiO2_SiC_Si} compares the MESM results with FEM simulations for the multilayer structure. Importantly, the spectrum of the combined SiO$_2$/SiC/Si structure is not simply a superposition of the responses of the individual SiO$_2$ and SiC layers. The resonant features shift and distort relative to those of the isolated layers, indicating electromagnetic coupling between the modes supported by the two materials. The MESM and FEM results agree well across the entire spectral range for both the amplitude and phase of the near-field response, including the frequency regions associated with the optical phonon resonances of both SiO$_2$ and SiC. This confirms that the generalized image-charge representation accurately describes structures with more than one finite-thickness layer.

\section{Application of MESM to uniaxial hyperbolic layers}

\begin{figure}
\centering
\includegraphics[width=\linewidth]{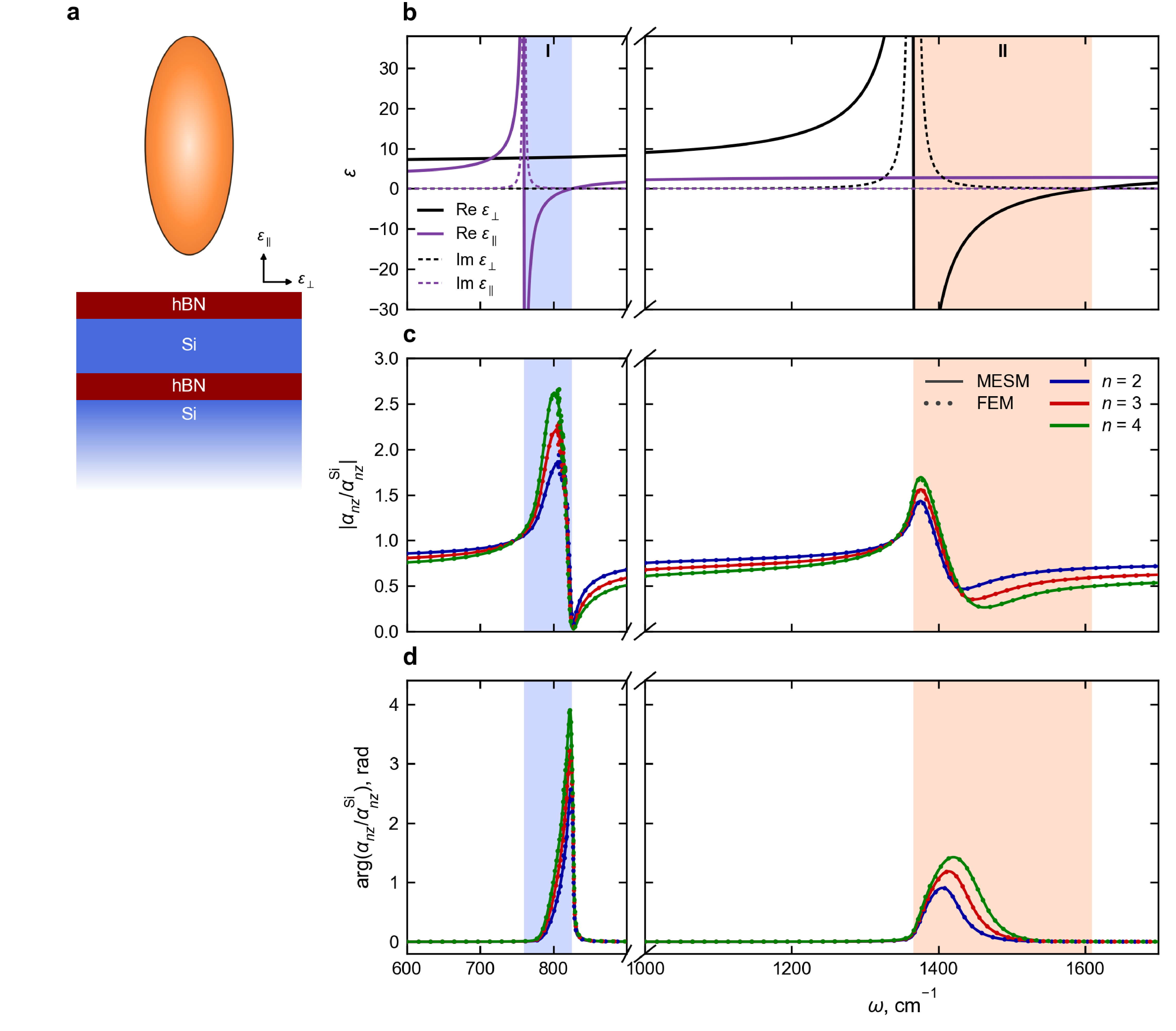}
\caption{
a) Schematic of the spheroid above a multilayer structure consisting of two 10-nm-thick hBN layers separated by a 20-nm-thick Si layer on a semi-infinite Si substrate. The black arrows indicate the out-of-plane, $\varepsilon_\parallel$, and in-plane, $\varepsilon_\perp$, directions.
b) Real (solid curves) and imaginary (dashed curves) parts of the in-plane, $\varepsilon_\perp$ (black), and out-of-plane, $\varepsilon_\parallel$ (purple), dielectric permittivities of hBN used in the calculations, Equation~\eqref{eq:S72}. The vertical axis is truncated. Shaded regions indicate the type-I (I) and type-II (II) hyperbolic bands.
c,d) Amplitude (c) and phase (d) of the second, third, and fourth harmonics of the spheroid polarizability above the sample, normalized to the corresponding response above bulk Si. Solid curves show the MESM results and dotted curves show the FEM simulations. The range 900--1000~cm$^{-1}$, for which no FEM simulations were performed, is omitted. The spheroid parameters are $R=30$~nm, $L=600$~nm, $A=50$~nm, and $H_{\min}=2$~nm.
}
\label{fig:S6}
\end{figure}

To demonstrate MESM's applicability to layers with strongly anisotropic optical responses, we consider a multilayer structure containing two layers of hexagonal boron nitride (hBN), a uniaxial van der Waals crystal whose optic axis is perpendicular to the layers \cite{caldwell2014,dai2014}. The sample consists of a 10-nm-thick hBN layer on a 20-nm-thick Si layer, followed by a second 10-nm-thick hBN layer on a semi-infinite Si substrate (Figure~\ref{fig:S6}a). The optic axes of both hBN layers are parallel to the spheroid axis, so that the system remains axially symmetric. The dielectric tensor of hBN is $\hat{\varepsilon}=\mathrm{diag} \left(\varepsilon_\perp, \varepsilon_\perp, \varepsilon_\parallel \right)$, where, as in Section~S2, $\varepsilon_\perp$ and $\varepsilon_\parallel$ denote the in-plane and out-of-plane dielectric permittivities, respectively. Both components are described by a single Lorentz oscillator,
\begin{equation}
\varepsilon_{\perp,\parallel}(\omega)=\varepsilon_\infty
\left[1+ \frac{\omega_{\mathrm{LO}}^2-\omega_{\mathrm{TO}}^2}{\omega_{\mathrm{TO}}^2-\omega^2-i\Gamma\omega} \right],
\label{eq:S72}
\end{equation}
with $\varepsilon_\infty=4.9$, $\omega_{\mathrm{TO}}=1366.2$~cm$^{-1}$, $\omega_{\mathrm{LO}}=1610$~cm$^{-1}$, and $\Gamma=7$~cm$^{-1}$ for $\varepsilon_\perp$, and $\varepsilon_\infty=2.95$, $\omega_{\mathrm{TO}}=760$~cm$^{-1}$, $\omega_{\mathrm{LO}}=825$~cm$^{-1}$, and $\Gamma=2$~cm$^{-1}$ for $\varepsilon_\parallel$ (Figure~\ref{fig:S6}b). Si is described by a frequency-independent dielectric permittivity $\varepsilon_{\mathrm{Si}}=12$. With these dielectric functions, hBN is hyperbolic in two spectral bands. In the type-I band between 760 and 825~cm$^{-1}$, $\mathrm{Re}\,\varepsilon_\parallel < 0 < \mathrm{Re}\,\varepsilon_\perp$, whereas in the type-II band between 1366 and 1610~cm$^{-1}$, $\mathrm{Re}\,\varepsilon_\perp < 0 < \mathrm{Re}\,\varepsilon_\parallel$. In both bands, hBN supports hyperbolic phonon polaritons \cite{caldwell2014,dai2014}. The spheroid parameters are $R=30$~nm, $L=600$~nm, $A=50$~nm, and $H_{\min}=2$~nm. All calculated spectra are normalized to the corresponding response above bulk Si.

Because hBN supports multiple polaritonic modes within the hyperbolic bands, its electrostatic reflection coefficient varies rapidly with $k_\parallel$, so the GICM representation converges slowly. We therefore evaluate the spheroid--sample coupling using FIM (Equation~\eqref{Sph_int_J_FIM}). For a uniaxial layer of thickness $d$ with the optic axis normal to the interfaces, the electrostatic reflection coefficient can be written in the same form as for an isotropic layer, Equations~\eqref{Re_layer} and \eqref{Beta_layer}, by replacing $\varepsilon_2$ with the effective permittivity $\bar{\varepsilon}=\sqrt{\varepsilon_\perp\varepsilon_\parallel}$, as in Equation~\eqref{Re_uniax}, and replacing $d$ with $\sqrt{\frac{\varepsilon_\perp}{\varepsilon_\parallel}} d$. For the complete multilayer structure, we obtain the electrostatic reflection coefficient recursively by applying the corresponding interface relations through the stack. We calculated the MESM spectra using the numerical tool described in Section~S11, with a multipole truncation order of $N=60$, $M=80$ sampling points over the tapping cycle, and a relative tolerance of $10^{-6}$ for the numerical integration over $k_\parallel$.

Figures~\ref{fig:S6}c,d compare the amplitude and phase of the second, third, and fourth harmonics of the spheroid polarizability obtained using MESM with the FEM simulations described in Section~S6. The FEM calculations were performed over the spectral ranges 600--900 and 1000--1700~cm$^{-1}$ with a frequency step of 1~cm$^{-1}$. The resulting spectra exhibit a pronounced resonance within each hyperbolic band. In the type-I band, a narrow resonance appears with amplitude maxima between 1.8 and 2.6 at frequencies of 801--806~cm$^{-1}$, depending on the demodulation harmonic. In the type-II band, a broader resonance appears close to the low-frequency edge of the band, with amplitude maxima between 1.4 and 1.7 at approximately 1376~cm$^{-1}$. In both cases, the resonance amplitude increases with increasing harmonic order. The MESM and FEM results agree well over the entire calculated spectral range, for both the amplitude and the phase of all three harmonics. Importantly, MESM reproduces the positions, amplitudes, and overall line shapes of the resonances in both hyperbolic bands. These results demonstrate that MESM, with the spheroid--sample coupling evaluated using FIM, accurately describes the near-field response of multilayer structures containing uniaxial hyperbolic layers.

\section{FDM spectra of SiC on non-excited and photoexcited InAs}

To further illustrate the limitations of the finite-dipole model (FDM) for the SiC/InAs system, we examine whether a single set of model parameters can describe both the non-excited and photoexcited states of the InAs substrate. All geometrical and material parameters are identical to those used in the main text; only the phenomenological complex-valued parameter $g$ is varied. We use the following dielectric permittivity of SiC:
\begin{equation}
\varepsilon_{\mathrm{SiC}}(\omega)=\varepsilon_{\infty,\mathrm{SiC}}
\left[1+ \frac{\omega_{\mathrm{LO}}^2-\omega_{\mathrm{TO}}^2}{\omega_{\mathrm{TO}}^2-\omega^2-i\Gamma\omega} \right],
\label{EpsSiC}
\end{equation}
with $\varepsilon_\infty=6.6$, $\omega_{\mathrm{TO}}=797$~cm$^{-1}$, $\omega_{\mathrm{LO}}=973$~cm$^{-1}$, and $\Gamma=6$~cm$^{-1}$. We describe the dielectric permittivity of photoexcited InAs using the Drude model:
\begin{equation}
\varepsilon_{\mathrm{InAs}}(\omega)=\varepsilon_{\infty,\mathrm{InAs}}
\left[1 - \frac{\omega_{\mathrm{p}}^2}{\omega^2+i\gamma_\mathrm{p}\omega} \right],
\label{EpsInAs}
\end{equation}
where $\varepsilon_{\infty,\mathrm{InAs}}=11.8$, $\omega_{\mathrm{p}}=845$~cm$^{-1}$, $\gamma_\mathrm{p}=125$~cm$^{-1}$. For non-excited InAs, we use $\varepsilon_{\mathrm{InAs}}(\omega)=\varepsilon_{\infty,\mathrm{InAs}}$.

\begin{figure}
\centering
\includegraphics[width=12cm]{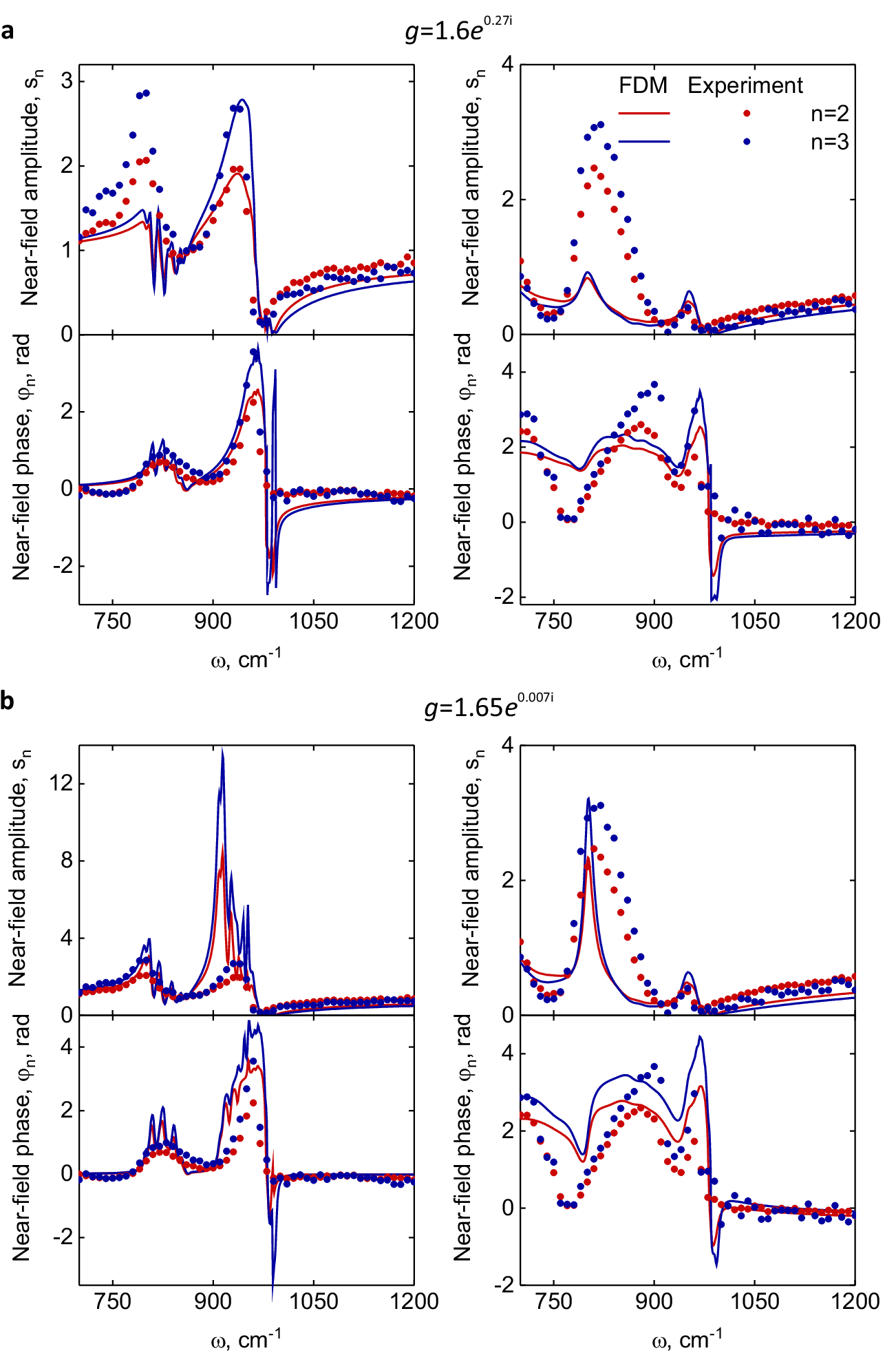}
\caption{
a) Experimental and FDM-calculated spectra of SiC on non-excited (left) and photoexcited (right) InAs using $g=1.6e^{0.27i}$, optimized for the non-excited substrate.
b) Corresponding spectra calculated using $g=1.65e^{0.007i}$, optimized for the photoexcited substrate. In both cases, all other model parameters are identical to those used in the main text.
}
\label{FDM}
\end{figure}

We first use $g=1.6e^{0.27i}$, which provides the best agreement with the experimental spectrum for SiC on non-excited InAs (Figure~\ref{FDM}a). In this case, FDM reproduces the main higher-frequency peak reasonably well. However, instead of reproducing the lower-frequency spectral feature, the calculated response exhibits rapid oscillations, indicating poor numerical convergence in this spectral region. When the same value of $g$ is applied to the photoexcited InAs substrate, the agreement deteriorates substantially: the main experimental resonance is not reproduced, and the calculated response remains nearly flat in the corresponding spectral region.

Conversely, the value $g=1.65e^{0.007i}$ provides a substantially better description of the photoexcited case (Figure~\ref{FDM}b). With this value, the overall spectral shape is reproduced qualitatively, while the amplitude of the main resonance is reproduced quantitatively. However, transferring the same parameter to the non-excited InAs substrate leads to a pronounced disagreement with experiment. In particular, the calculated spectral peak is several times larger than the measured one, and narrow spurious spikes appear near both resonant features.

These results show that, although the FDM can be independently adjusted to reproduce important features of either the non-excited or photoexcited spectra, no single value of the $g$-parameter consistently describes both optical states of the same physical SiC/InAs structure. This contrasts with the MESM results presented in the main text, where the same geometrical parameters describe both states without readjustment.

\section{Publicly available numerical tool}

To make the MESM accessible without programming, we implemented it in an open-source application, the s-SNOM Spectrum Calculator, written in Python with the Streamlit framework. The application calculates the normalized near-field amplitude \(s_n\) and phase \(\varphi_n\) for a bulk sample or for an ordered stack of finite layers on a semi-infinite substrate, with the tip modeled as a prolate spheroid. The harmonics \(n=0,1,2,3,4,5\) can be calculated and visualised. The calculation method is chosen automatically from the sample geometry and the symmetry of the permittivity tensors: isotropic layers are treated with GICM (Sections S2 and S3), and uniaxial layers with FIM (Sections S4 and S5), as in Section S9. For isotropic layers, the FIM is also used automatically for very strong oscillators, where the GICM would require more than 16 image charges and the FIM becomes slightly faster (Section S7). It can also be selected explicitly to cross-check the GICM result. The source code is available in a GitHub repository \cite{snom_calculator_github} and archived on Zenodo with a permanent DOI \cite{snom_calculator_zenodo}. A hosted version of the application runs in a web browser at \url{https://snomcalculator.streamlit.app/}, so that spectra can be calculated without installing any software.

\begin{figure}[htbp]
\centering
\includegraphics[width=\textwidth]{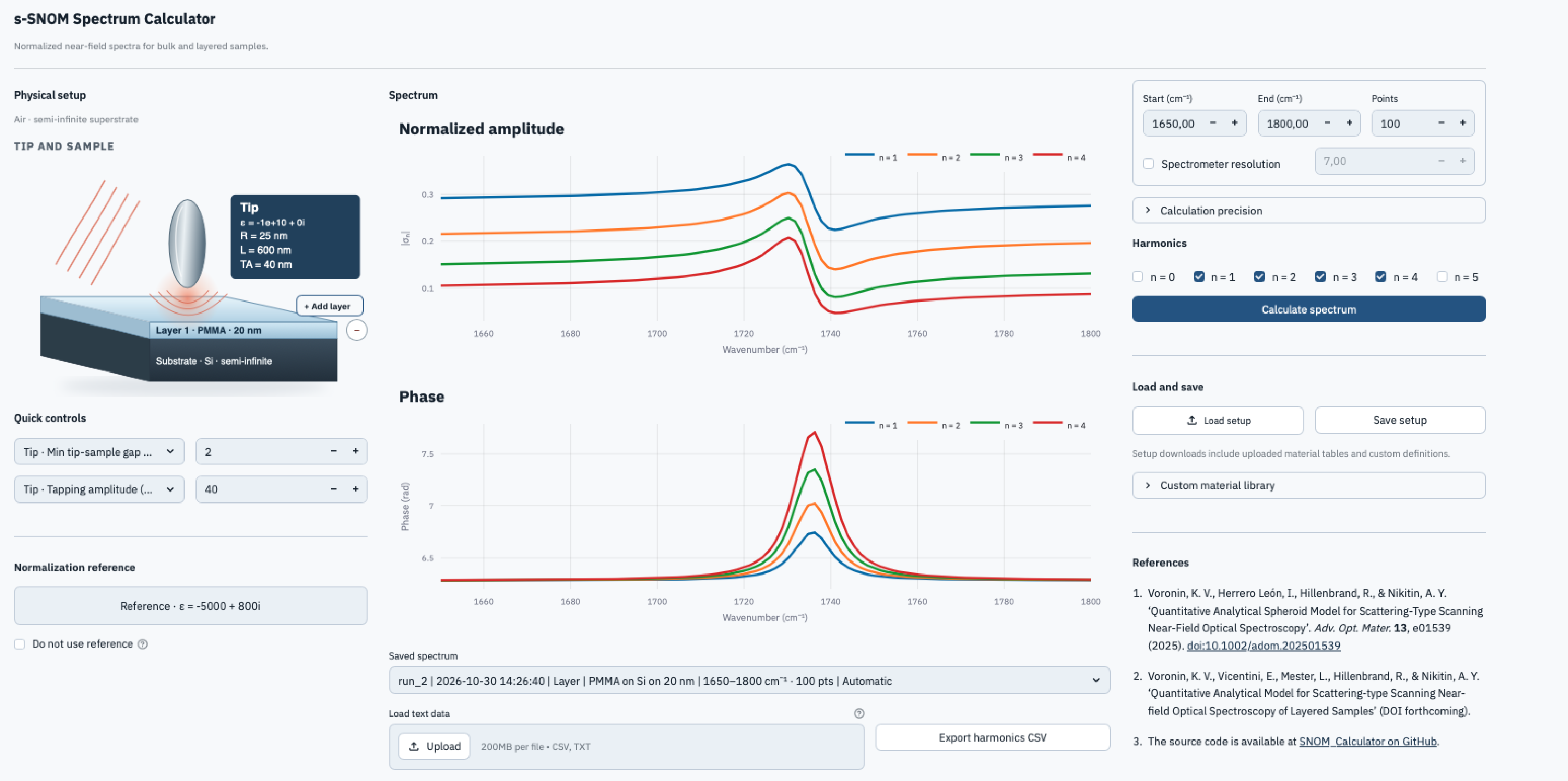}
\caption{Screenshot of the s-SNOM Spectrum Calculator. The left column shows the physical setup, namely the tip, a finite layer, and the substrate, together with the quick controls and the normalization reference. The central panel shows the normalized amplitude and phase of the first to fourth harmonics, \(n=1\)--4, as functions of the wavenumber, calculated for a 20-nm-thick PMMA layer on a semi-infinite silicon substrate (Si in the catalog). The right column contains the frequency range, the numerical settings, the selection of the harmonics, and the controls for loading and saving a setup. The spheroid parameters are \(R=25\) nm, \(L=600\) nm, \(A=40\) nm, and \(H_{\min}=2\) nm, and the spectra are normalized to the response of a reference with \(\varepsilon=-5000+800i\).}
\label{fig:S8}
\end{figure}

The user builds the sample in a single browser window (Figure~\ref{fig:S8}): the tip parameters, the layers, the substrate, and the normalization reference are set in a graphical editor. Materials are taken from the built-in catalog \cite{materials_library}, entered as a constant permittivity or a tabulated \(\varepsilon(\omega)\), or defined from a Drude term and up to six Lorentz oscillators. Materials defined in this way can be collected in a personal library, which the user can edit, export, and import, and which is saved automatically in the local installation. Calculated spectra can be compared with imported experimental data and exported together with the calculation settings, and a complete setup can be saved and restored. The same engines can also be called from Python.

The numerical parameters \(N\), \(M\) and  \(K\) are set automatically from the strength of the material oscillators, the type of sample, and its distance from the tip, using empirical formulas  obtained from convergence calculations for more than 500 different configurations. They are designed to keep the error of the demodulated spectra below 5\% in at least 95\% of the tested configurations; the remaining cases occur mostly for strong and very strong oscillators at very small tip--sample gaps (less than 2~nm). For each calculation, the application compares \(M\) with \(M/2\) and \(N\) with \(N/2\), and offers to double a parameter when the difference exceeds 20\%. This threshold is a conservative indicator for achieving a final error below 5\% along all calculated the spectra, not a certified error bound, and users should verify convergence themselves, especially for small gaps, long tips, and sharp resonances.

The computation time increases with the number of frequency points, with the required precision, and with the complexity of the sample, in particular for uniaxial layers. The hosted version runs on the limited, shared computing resources of Streamlit Cloud, so extensive calculations can be slow there. For such calculations, we recommend installing the application locally, which uses the full resources of the user's own computer and therefore runs faster. After downloading the repository from GitHub, the user runs the installation launcher for Windows or macOS, which prepares a local Python environment and opens the application in the browser; no prior knowledge of Python is needed. A step-by-step installation guide, the user documentation, and the description of the numerical models are provided in the repository. We kindly ask users to cite the main text and the archived version of the software \cite{snom_calculator_zenodo} when the application is used in published work.

\sloppy\emergencystretch=3em
\bibliography{SIRefs}